\documentclass[longauth]{aa} 
\usepackage{graphicx}
\usepackage{txfonts}
\usepackage{bigints}
\usepackage{enumitem}
\usepackage[breaklinks=true]{hyperref} 
\usepackage{wasysym}

\def\gaia{{\em Gaia}}

\def\parallax{$\varpi$}
\def\teff{$T_{\rm eff}$}
\def\logg{$\log g$}

\def\mh{[M/H]}
\def\vsini{$v\sin i$}

\def\ha10{${\rm H}\alpha\,10\%$}

\def\espcs{ESP-CS}
\def\gspspec{GSP-Spec}
\def\gspphot{GSP-Phot}

\def\bmr{$(G_{\rm BP}-G_{\rm RP})$}
\def\mg{$M_G$}
\def\rprimeHK{$R'_{\rm{HK}}$}

\providecommand{\teff}{\ensuremath{T_{\mathrm{{eff}}}}\xspace}

\providecommand{\logg}{\ensuremath{\log\,g}\xspace}

\providecommand{\mh}{[M/H]\xspace}

\providecommand{\vsini}{\ensuremath{v\sin i}\xspace}

\providecommand{\gmag}{\ensuremath{G}}

\providecommand{\mg}{M$_\gmag$}

\providecommand{\parallax}{\ensuremath{\varpi}}

\providecommand{\nm}{\ensuremath{\,\mathrm{nm}}\xspace}

\providecommand{\gspphot}{\modulename{GSP-Phot}}

\providecommand{\gspspec}{\modulename{GSP-Spec}}

\providecommand{\espcs}{\modulename{ESP-CS}}

\providecommand\gaia{\textit{Gaia}}
\providecommand{\gdr}[1]{Gaia~DR{#1}}

\providecommand{\linktodoc}{https://geapre.esac.esa.int/archive/documentation/GDR3}

\providecommand{\linkfig}[1]{\href{\linktodoc/Data_analysis/chap_cu8par/#1}{see table\xspace}}

\makeatletter
\DeclareRobustCommand*{\fieldName}[1]{%
  \begingroup\@fieldName\scantokens{\texttt{\small {#1}}\noexpand}\endgroup}
\begingroup\lccode`\~=`\_\relax
   \lowercase{\endgroup\def\@fieldName{\catcode`\_=\active \let~\_}}
\makeatother

\begin{document}

   \title{{\it Gaia} Data Release 3}

   \subtitle{Stellar chromospheric activity and mass accretion from \ion{Ca}{ii} IRT observed by the Radial Velocity Spectrometer}

	\titlerunning{{\it Gaia} DR3: Stellar activity}


\authorrunning{A. C. Lanzafame et al.}

\author{
A.C.~                     Lanzafame\orcid{0000-0002-2697-3607}\inst{\ref{inst:0001},\ref{inst:0002}}
\and         E.~                    Brugaletta\orcid{0000-0003-2598-6737}\inst{\ref{inst:0001}}
\and         Y.~                    Fr\'{e}mat\orcid{0000-0002-4645-6017}\inst{\ref{inst:0004}}
\and         R.~                         Sordo\orcid{0000-0003-4979-0659}\inst{\ref{inst:0005}}
\and       O.L.~                       Creevey\orcid{0000-0003-1853-6631}\inst{\ref{inst:0006}}
\and         V.~                       Andretta\orcid{0000-0003-1962-9741}\inst{\ref{inst:oana}}
\and         G.~                    Scandariato\orcid{0000-0003-2029-0626}\inst{\ref{inst:0001}}
\and         I.~                         Bus\`a\orcid{0000-0003-2876-3563}\inst{\ref{inst:0001}}
\and         E.~.                     Distefano\orcid{0000-0002-2448-2513}\inst{\ref{inst:0001}}
\and       A.J.~                          Korn\orcid{0000-0002-3881-6756}\inst{\ref{inst:0007}}
\and         P.~                    de Laverny\orcid{0000-0002-2817-4104}\inst{\ref{inst:0006}}
\and         A.~                  Recio-Blanco\orcid{0000-0002-6550-7377}\inst{\ref{inst:0006}}
\and         A.~                Abreu Aramburu\inst{\ref{inst:0010}}
\and       M.A.~                   \'{A}lvarez\orcid{0000-0002-6786-2620}\inst{\ref{inst:0011}}
\and         R.~                        Andrae\orcid{0000-0001-8006-6365}\inst{\ref{inst:0012}}
\and     C.A.L.~                  Bailer-Jones\inst{\ref{inst:0012}}
\and         J.~                        Bakker\inst{\ref{inst:0035}}
\and         I.~                Bellas-Velidis\inst{\ref{inst:0014}}
\and         A.~                       Bijaoui\inst{\ref{inst:0006}}
\and         N.~                     Brouillet\orcid{0000-0002-3274-7024}\inst{\ref{inst:0016}}
\and         A.~                       Burlacu\inst{\ref{inst:0017}}
\and         R.~                      Carballo\orcid{0000-0001-7412-2498}\inst{\ref{inst:0018}}
\and         L.~                   Casamiquela\orcid{0000-0001-5238-8674}\inst{\ref{inst:0016},\ref{inst:0020}}
\and         L.~                        Chaoul\inst{\ref{inst:0021}}
\and         A.~                     Chiavassa\orcid{0000-0003-3891-7554}\inst{\ref{inst:0006}}
\and         G.~                      Contursi\orcid{0000-0001-5370-1511}\inst{\ref{inst:0006}}
\and       W.J.~                        Cooper\orcid{0000-0003-3501-8967}\inst{\ref{inst:0024},\ref{inst:0025}}
\and         C.~                       Dafonte\orcid{0000-0003-4693-7555}\inst{\ref{inst:0011}}
\and         A.~                    Dapergolas\inst{\ref{inst:0014}}
\and         L.~                    Delchambre\orcid{0000-0003-2559-408X}\inst{\ref{inst:0028}}
\and         C.~                      Demouchy\inst{\ref{inst:0029}}
\and       T.E.~                 Dharmawardena\orcid{0000-0002-9583-5216}\inst{\ref{inst:0012}}
\and         R.~                       Drimmel\orcid{0000-0002-1777-5502}\inst{\ref{inst:0025}}
\and         B.~                    Edvardsson\inst{\ref{inst:0032}}
\and         M.~                     Fouesneau\orcid{0000-0001-9256-5516}\inst{\ref{inst:0012}}
\and         D.~                      Garabato\orcid{0000-0002-7133-6623}\inst{\ref{inst:0011}}
\and         P.~              Garc\'{i}a-Lario\orcid{0000-0003-4039-8212}\inst{\ref{inst:0035}}
\and         M.~             Garc\'{i}a-Torres\orcid{0000-0002-6867-7080}\inst{\ref{inst:0036}}
\and         A.~                         Gavel\orcid{0000-0002-2963-722X}\inst{\ref{inst:0007}}
\and         A.~                         Gomez\orcid{0000-0002-3796-3690}\inst{\ref{inst:0011}}
\and         I.~   Gonz\'{a}lez-Santamar\'{i}a\orcid{0000-0002-8537-9384}\inst{\ref{inst:0011}}
\and         D.~                Hatzidimitriou\orcid{0000-0002-5415-0464}\inst{\ref{inst:0040},\ref{inst:0014}}
\and         U.~                        Heiter\orcid{0000-0001-6825-1066}\inst{\ref{inst:0007}}
\and         A.~          Jean-Antoine Piccolo\orcid{0000-0001-8622-212X}\inst{\ref{inst:0021}}
\and         M.~                      Kontizas\orcid{0000-0001-7177-0158}\inst{\ref{inst:0040}}
\and         G.~                    Kordopatis\orcid{0000-0002-9035-3920}\inst{\ref{inst:0006}}
\and         Y.~                      Lebreton\orcid{0000-0002-4834-2144}\inst{\ref{inst:0046},\ref{inst:0047}}
\and       E.L.~                        Licata\orcid{0000-0002-5203-0135}\inst{\ref{inst:0025}}
\and     H.E.P.~                  Lindstr{\o}m\inst{\ref{inst:0025},\ref{inst:0050},\ref{inst:0051}}
\and         E.~                       Livanou\orcid{0000-0003-0628-2347}\inst{\ref{inst:0040}}
\and         A.~                         Lobel\orcid{0000-0001-5030-019X}\inst{\ref{inst:0004}}
\and         A.~                         Lorca\inst{\ref{inst:0054}}
\and         A.~               Magdaleno Romeo\inst{\ref{inst:0017}}
\and         M.~                      Manteiga\orcid{0000-0002-7711-5581}\inst{\ref{inst:0056}}
\and         F.~                       Marocco\orcid{0000-0001-7519-1700}\inst{\ref{inst:0057}}
\and       D.J.~                      Marshall\orcid{0000-0003-3956-3524}\inst{\ref{inst:0058}}
\and         N.~                          Mary\inst{\ref{inst:0059}}
\and         C.~                       Nicolas\inst{\ref{inst:0021}}
\and         C.~                     Ordenovic\inst{\ref{inst:0006}}
\and         F.~                       Pailler\orcid{0000-0002-4834-481X}\inst{\ref{inst:0021}}
\and       P.A.~                       Palicio\orcid{0000-0002-7432-8709}\inst{\ref{inst:0006}}
\and         L.~               Pallas-Quintela\orcid{0000-0001-9296-3100}\inst{\ref{inst:0011}}
\and         C.~                         Panem\inst{\ref{inst:0021}}
\and         B.~                        Pichon\orcid{0000 0000 0062 1449}\inst{\ref{inst:0006}}
\and         E.~                        Poggio\orcid{0000-0003-3793-8505}\inst{\ref{inst:0006},\ref{inst:0025}}
\and         F.~                        Riclet\inst{\ref{inst:0021}}
\and         C.~                         Robin\inst{\ref{inst:0059}}
\and         J.~                       Rybizki\orcid{0000-0002-0993-6089}\inst{\ref{inst:0012}}
\and         R.~                 Santove\~{n}a\orcid{0000-0002-9257-2131}\inst{\ref{inst:0011}}
\and       L.M.~                         Sarro\orcid{0000-0002-5622-5191}\inst{\ref{inst:0073}}
\and       M.S.~                    Schultheis\orcid{0000-0002-6590-1657}\inst{\ref{inst:0006}}
\and         M.~                         Segol\inst{\ref{inst:0029}}
\and         A.~                       Silvelo\orcid{0000-0002-5126-6365}\inst{\ref{inst:0011}}
\and         I.~                        Slezak\inst{\ref{inst:0006}}
\and       R.L.~                         Smart\orcid{0000-0002-4424-4766}\inst{\ref{inst:0025}}
\and         C.~                      Soubiran\orcid{0000-0003-3304-8134}\inst{\ref{inst:0016}}
\and         M.~                  S\"{ u}veges\orcid{0000-0003-3017-5322}\inst{\ref{inst:0080}}
\and         F.~                  Th\'{e}venin\inst{\ref{inst:0006}}
\and         G.~                Torralba Elipe\orcid{0000-0001-8738-194X}\inst{\ref{inst:0011}}
\and         A.~                          Ulla\orcid{0000-0001-6424-5005}\inst{\ref{inst:0083}}
\and.        E.~                       Utrilla\inst{\ref{inst:0054}}
\and         A.~                     Vallenari\orcid{0000-0003-0014-519X}\inst{\ref{inst:0005}}
\and         E.~                    van Dillen\inst{\ref{inst:0029}}
\and         H.~                          Zhao\orcid{0000-0003-2645-6869}\inst{\ref{inst:0006}}
\and         J.~                         Zorec\inst{\ref{inst:0087}}
}
\institute{
     INAF - Osservatorio Astrofisico di Catania, via S. Sofia 78, 95123 Catania, Italy\relax                                                                                                                                                                                                                                                                       \label{inst:0001}
\and Dipartimento di Fisica e Astronomia ""Ettore Majorana"", Universit\`{a} di Catania, Via S. Sofia 64, 95123 Catania, Italy\relax                                                                                                                                                                                                                               \label{inst:0002}\vfill
\and Royal Observatory of Belgium, Ringlaan 3, 1180 Brussels, Belgium\relax                                                                                                                                                                                                                                                                                        \label{inst:0004}\vfill
\and INAF - Osservatorio astronomico di Padova, Vicolo Osservatorio 5, 35122 Padova, Italy\relax                                                                                                                                                                                                                                                                   \label{inst:0005}\vfill
\and Universit\'{e} C\^{o}te d'Azur, Observatoire de la C\^{o}te d'Azur, CNRS, Laboratoire Lagrange, Bd de l'Observatoire, CS 34229, 06304 Nice Cedex 4, France\relax                                                                                                                                                                                              \label{inst:0006}\vfill
\and INAF- Osservatorio Astronomico di Capodimonte, Salita Moiariello 16, 80131 Naples, Italy\relax\label{inst:oana}\vfill
\and Observational Astrophysics, Division of Astronomy and Space Physics, Department of Physics and Astronomy, Uppsala University, Box 516, 751 20 Uppsala, Sweden\relax                                                                                                                                                                                           \label{inst:0007}\vfill
\and ATG Europe for European Space Agency (ESA), Camino bajo del Castillo, s/n, Urbanizacion Villafranca del Castillo, Villanueva de la Ca\~{n}ada, 28692 Madrid, Spain\relax                                                                                                                                                                                      \label{inst:0010}\vfill
\and CIGUS CITIC - Department of Computer Science and Information Technologies, University of A Coru\~{n}a, Campus de Elvi\~{n}a s/n, A Coru\~{n}a, 15071, Spain\relax                                                                                                                                                                                             \label{inst:0011}\vfill
\and Max Planck Institute for Astronomy, K\"{ o}nigstuhl 17, 69117 Heidelberg, Germany\relax                                                                                                                                                                                                                                                                       \label{inst:0012}\vfill
\and National Observatory of Athens, I. Metaxa and Vas. Pavlou, Palaia Penteli, 15236 Athens, Greece\relax                                                                                                                                                                                                                                                         \label{inst:0014}\vfill
\and Laboratoire d'astrophysique de Bordeaux, Univ. Bordeaux, CNRS, B18N, all{\'e}e Geoffroy Saint-Hilaire, 33615 Pessac, France\relax                                                                                                                                                                                                                             \label{inst:0016}\vfill
\and Telespazio for CNES Centre Spatial de Toulouse, 18 avenue Edouard Belin, 31401 Toulouse Cedex 9, France\relax                                                                                                                                                                                                                                                 \label{inst:0017}\vfill
\and Dpto. de Matem\'{a}tica Aplicada y Ciencias de la Computaci\'{o}n, Univ. de Cantabria, ETS Ingenieros de Caminos, Canales y Puertos, Avda. de los Castros s/n, 39005 Santander, Spain\relax                                                                                                                                                                   \label{inst:0018}\vfill
\and GEPI, Observatoire de Paris, Universit\'{e} PSL, CNRS, 5 Place Jules Janssen, 92190 Meudon, France\relax                                                                                                                                                                                                                                                      \label{inst:0020}\vfill
\and CNES Centre Spatial de Toulouse, 18 avenue Edouard Belin, 31401 Toulouse Cedex 9, France\relax                                                                                                                                                                                                                                                                \label{inst:0021}\vfill
\and Centre for Astrophysics Research, University of Hertfordshire, College Lane, AL10 9AB, Hatfield, United Kingdom\relax                                                                                                                                                                                                                                         \label{inst:0024}\vfill
\and INAF - Osservatorio Astrofisico di Torino, via Osservatorio 20, 10025 Pino Torinese (TO), Italy\relax                                                                                                                                                                                                                                                         \label{inst:0025}\vfill
\and Institut d'Astrophysique et de G\'{e}ophysique, Universit\'{e} de Li\`{e}ge, 19c, All\'{e}e du 6 Ao\^{u}t, B-4000 Li\`{e}ge, Belgium\relax                                                                                                                                                                                                                    \label{inst:0028}\vfill
\and APAVE SUDEUROPE SAS for CNES Centre Spatial de Toulouse, 18 avenue Edouard Belin, 31401 Toulouse Cedex 9, France\relax                                                                                                                                                                                                                                        \label{inst:0029}\vfill
\and Theoretical Astrophysics, Division of Astronomy and Space Physics, Department of Physics and Astronomy, Uppsala University, Box 516, 751 20 Uppsala, Sweden\relax                                                                                                                                                                                             \label{inst:0032}\vfill
\and European Space Agency (ESA), European Space Astronomy Centre (ESAC), Camino bajo del Castillo, s/n, Urbanizacion Villafranca del Castillo, Villanueva de la Ca\~{n}ada, 28692 Madrid, Spain\relax                                                                                                                                                             \label{inst:0035}\vfill
\and Data Science and Big Data Lab, Pablo de Olavide University, 41013, Seville, Spain\relax                                                                                                                                                                                                                                                                       \label{inst:0036}\vfill
\and Department of Astrophysics, Astronomy and Mechanics, National and Kapodistrian University of Athens, Panepistimiopolis, Zografos, 15783 Athens, Greece\relax                                                                                                                                                                                                  \label{inst:0040}\vfill
\and LESIA, Observatoire de Paris, Universit\'{e} PSL, CNRS, Sorbonne Universit\'{e}, Universit\'{e} de Paris, 5 Place Jules Janssen, 92190 Meudon, France\relax                                                                                                                                                                                                   \label{inst:0046}\vfill
\and Universit\'{e} Rennes, CNRS, IPR (Institut de Physique de Rennes) - UMR 6251, 35000 Rennes, France\relax                                                                                                                                                                                                                                                      \label{inst:0047}\vfill
\and Niels Bohr Institute, University of Copenhagen, Juliane Maries Vej 30, 2100 Copenhagen {\O}, Denmark\relax                                                                                                                                                                                                                                                    \label{inst:0050}\vfill
\and DXC Technology, Retortvej 8, 2500 Valby, Denmark\relax                                                                                                                                                                                                                                                                                                        \label{inst:0051}\vfill
\and Aurora Technology for European Space Agency (ESA), Camino bajo del Castillo, s/n, Urbanizacion Villafranca del Castillo, Villanueva de la Ca\~{n}ada, 28692 Madrid, Spain\relax                                                                                                                                                                               \label{inst:0054}\vfill
\and CIGUS CITIC, Department of Nautical Sciences and Marine Engineering, University of A Coru\~{n}a, Paseo de Ronda 51, 15071, A Coru\~{n}a, Spain\relax                                                                                                                                                                                                          \label{inst:0056}\vfill
\and IPAC, Mail Code 100-22, California Institute of Technology, 1200 E. California Blvd., Pasadena, CA 91125, USA\relax                                                                                                                                                                                                                                           \label{inst:0057}\vfill
\and IRAP, Universit\'{e} de Toulouse, CNRS, UPS, CNES, 9 Av. colonel Roche, BP 44346, 31028 Toulouse Cedex 4, France\relax                                                                                                                                                                                                                                        \label{inst:0058}\vfill
\and Thales Services for CNES Centre Spatial de Toulouse, 18 avenue Edouard Belin, 31401 Toulouse Cedex 9, France\relax                                                                                                                                                                                                                                            \label{inst:0059}\vfill
\and Dpto. de Inteligencia Artificial, UNED, c/ Juan del Rosal 16, 28040 Madrid, Spain\relax                                                                                                                                                                                                                                                                       \label{inst:0073}\vfill
\and Institute of Global Health, University of Geneva\relax                                                                                                                                                                                                                                                                                                        \label{inst:0080}\vfill
\and Applied Physics Department, Universidade de Vigo, 36310 Vigo, Spain\relax                                                                                                                                                                                                                                                                                     \label{inst:0083}\vfill
\and Sorbonne Universit\'{e}, CNRS, UMR7095, Institut d'Astrophysique de Paris, 98bis bd. Arago, 75014 Paris, France\relax                                                                                                                                                                                                                                         \label{inst:0087}\vfill
}

   \date{Received Month day, year; accepted Month day, year}

 
  \abstract
   {The \gaia\ Radial Velocity Spectrometer (RVS) provides the unique opportunity of a spectroscopic analysis of millions of stars at medium-resolution ($\lambda/\Delta \lambda \sim $11500) in the near-infrared (845 -- 872\,nm).
    This wavelength range includes the \ion{Ca}{ii} infrared triplet (IRT) at 850.03, 854.44, and 866.45\,nm, which is a good diagnostics of magnetic activity in the chromosphere of late-type stars. 
}
   {Here we present the method devised for inferring the \gaia\ stellar activity index from the analysis of the \ion{Ca}{ii} IRT in the RVS spectrum, together with its scientific validation.
   }
   {The \gaia\ stellar activity index is derived from the \ion{Ca}{ii} IRT excess equivalent width with respect to a reference spectrum, taking the projected rotational velocity (\vsini) into account. 
   Scientific validation of the \gaia\ stellar activity index 
   is performed by deriving a $R'_{\rm IRT}$ index, largely independent of the photospheric parameters, and considering the correlation with the $R'_{\rm HK}$ index for a sample of stars.
   A sample of well studied PMS stars is considered to identify the regime in which the \gaia\ stellar activity index may be affected by mass accretion.
   The position of these stars in the colour-magnitude diagram and the correlation with the amplitude of the photometric rotational modulation is also scrutinised.
	}
   {
   \gdr{3} contains a stellar activity index derived from the \ion{Ca}{ii} IRT for some $2\times10^6$ stars in the Galaxy.
   This represents a gold mine for studies on stellar magnetic activity and mass accretion in the solar vicinity.
   Three regimes of the chromospheric stellar activity are identified, confirming suggestions made by previous authors on much smaller $R'_{\rm HK}$ datasets.
   The highest stellar activity regime is associated with PMS stars and RS\,CVn systems, in which activity is enhanced by tidal interaction.
   Some evidence of a bimodal distribution in MS stars with \teff$\apprge$\,5000\,K is also found, which defines the two other regimes, without a clear gap in between.
   Stars with 3500\,K$\apprle$\teff$\apprle$\,5000\,K are found to be either very active PMS stars or active MS stars with a unimodal distribution in chromospheric activity.
   A dramatic change in the activity distribution is found for \teff$\apprle$3500\,K, with a dominance of low activity stars close to the transition between partially- and fully-convective stars and a rise in activity down into the fully-convective regime.
   }
   {}


   \maketitle
%

\section{Introduction}

A stellar chromosphere is the manifestation of non-radiative heating due to (magneto)acoustic waves dissipation and magnetic-field recombination in late-type stellar atmosphere
\citep[see ][ for recent reviews]{2017ARA&A..55..159L,2019ARA&A..57..189C}.
Such a non-radiative heating becomes increasingly important in the energy balance at increasing heights (decreasing gas density) above the stellar photosphere, and leads to a temperature increase above the radiative equilibrium value, as revealed by variable emission in X-rays, UV, radio, and in some optical spectral lines like the \ion{Ca}{ii}\,H\&K doublet, the H$\alpha$ and the \ion{Ca}{ii} infrared triplet (IRT) at $\lambda$\,850.03, 854.44, and 866.45\,nm (vacuum).
A large fraction of the stellar non-radiative energy is dissipated in the chromosphere while the remaining is transferred to the outer atmosphere producing a variety of phenomena including the heating of the corona and the stellar wind.

The magnetic fields in the stellar outer atmosphere of late-type stars  emerge from the stellar interior where they are generated by a complex dynamo mechanisms involving turbulence and rotation.
The angular momentum loss via the magnetised wind causes the star to spin-down, so that the dynamo efficiency decreases in time.
As a consequence, many phenomena connected to the surface magnetic field decay as the star ages.
These include surface rotation, photospheric spots, chromospheric and coronal emission, as well as flares occurrence and energy.
The time decay of rotation and chromospheric activity in the early main sequence, in particular, are quite predictable and can be used as stellar clocks \citep[see, e.g. ][and references therein]{,2015A&A...584A..30L,2008ApJ...687.1264M}.
On the other hand, young, fast-rotating stars are in a ``saturated'' regime in which the correlation of magnetic activity with rotation is weak and our knowledge of the rotation evolution itself is still poor, which prevents using phenomena related to magnetic activity as stellar clocks in this regime \citep[see, e.g.,][and references therein]{2019ApJ...877..157L}.
For main sequence stars older than the Sun, the comparison with asteroseismology has spurred some controversy on the reliability of gyrochronology \citep{2003ApJ...586..464B,2010ApJ...722..222B,2015A&A...584A..30L,2020A&A...636A..76S} at low rotation rates, where wind-braking seems to become ineffective \citep{2016Natur.529..181V}. 
However, the method hinges on the knowledge of the metallicity \citep{2020MNRAS.499.3481A}, which is becoming available on a large scale only with recent spectroscopic surveys.
Note also that chromospheric activity is superimposed on a basal emission originating from the dissipation of acoustic energy \citep{1989ApJ...337..964S}, and that an investigation on the lowest level of chromospheric activity may help in understanding the general role of purely acoustic energy in the stellar chromosphere energy balance.

In pre-main sequence stars still accreting mass through the protostellar disk, emission induced by accretion shocks and magnetosphere is superimposed to the chromospheric emission, sometimes dominating completely over the chromospheric component \citep[see, e.g., ][]{2016ARA&A..54..135H,2017ARA&A..55..159L}.
Large surveys permit to establish empirically ``dividing lines'' between the regimes dominated by chromospheric activity or accretion \citep[e.g. ][]{2015A&A...576A..80L,2017ApJ...835...61Z}. 

Stellar magnetic activity plays a crucial role in the search of exoplanets, particularly of Earth-like planets.
The presence of spots, convective turbulence, and granulation induces a radial velocity (RV) ``jitter'' that may sometimes mimic the signal produced by the planets or hinder the signal produced by small planets.
For Sun-like stars, measurements of chromospheric emission in the \ion{Ca}{ii} H\& K lines, such as $R'_{\rm HK}$ \citep{1984ApJ...279..763N,1991ApJS...76..383D} have been shown to correlate with intrinsic stellar RV variations \citep{1988ApJ...331..902C, 1998ApJ...498L.153S,2000A&A...361..265S, 2005PASP..117..657W}. 
In these cases, the chromospheric magnetic activity is also correlated with the presence of photospheric spots or faculae, so that it turns out to be a useful proxy of the processes that produce the intrinsic RV jitter.
In the general case, the dependence on other stellar parameters has to be taken into account too, particularly the dependence on \logg\ \citep{2014AJ....147...29B} and therefore on the evolutionary stage of the star \citep[see also][and references therein]{2020AJ....159..235L}.
Therefore, measurements of the chromospheric magnetic activity, together with the fundamental stellar parameters, provide an efficient tool for understanding which type of stars presents poor cases for RV follow-up for planet hunting due to high jitter induced by magnetic activity. 

Stellar magnetic activity phenomena include  the atmospheric space weather, i.e. the perturbation traveling from stars to planets in the form of flares, winds, coronal mass ejections and energetic particles.
These phenomena may profoundly affect the dynamics, chemistry, and climate of exoplanets \citep[see, e.g., ][and references therein]{2020IJAsB..19..136A}, and, in some cases, even the retention of the exoplanetary atmosphere itself \citep[e.g. ][]{2020A&A...639A..49G}.

Magnetic activity has also an impact in the derivation of accurate elemental abundance in the stellar atmosphere.
Lines formed up in the photosphere are found to modulate over the stellar activity cycle \citep{2016A&A...589A.135F,2019MNRAS.490L..86Y}. 
Analysing HARPS high resolution spectra of 211 sunlike stars observed at different phases of their activity cycle, \cite{2020ApJ...895...52S} observed an increase in the equivalent width of the lines as a function of the activity index \rprimeHK\ (see Appendix \ref{sec:activity_indices}).
This effect is visible for stars with $\log$\,\rprimeHK$\ge-5.0$, increases with increasing activity, and produces an artificial growth of the stellar microturbulence and a decrease in effective temperature and metallicity.

As part of the activities carried out by the Coordination Unit 8 (CU8) of the Gaia Data Processing and Analysis Consortium (DPAC), we have implemented a method for deriving a stellar activity index using the \ion{Ca}{ii}\,IRT lines observed by the \gaia\ Radial Velocity Spectrometer (RVS).
The method is part of the Extended Stellar Parametrizer for Cool Stars, (\espcs) module of the  Astrophysical parameters inference system
(Apsis) \citep{2013A&A...559A..74B,DR3-DPACP-157,DR3-DPACP-160,onlinedocdr3}. 
In this paper we present this method together with the scientific validation of the results available in \gdr{3}.
The scientific validation, in particular, provides criteria for discriminating between the purely chromospheric activity regime and the mass accretion regime in T\,Tauri stars. 
In Sect.\,\ref{sec:method} we describe the method implemented, in Sect.\,\ref{sec:results} we give an outline of the results obtained together with the scientific validation of the data available in \gdr{3}, and in Sect.\,\ref{sec:conclusions} we draw our conclusions.

\section{Method}\label{sec:method}

\subsection{Input data}\label{sec:data}

\begin{figure*}[ht]
\begin{center}
\includegraphics[width=0.45\textwidth]{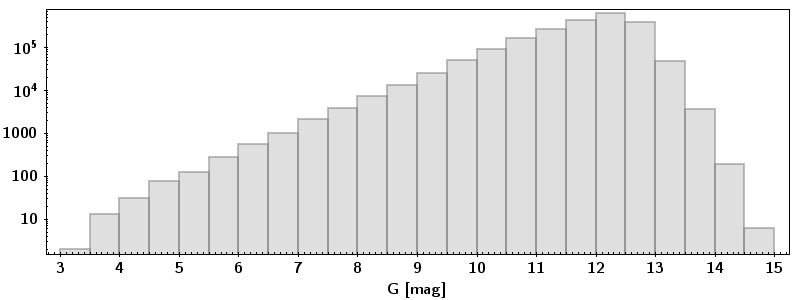}
\includegraphics[width=0.45\textwidth]{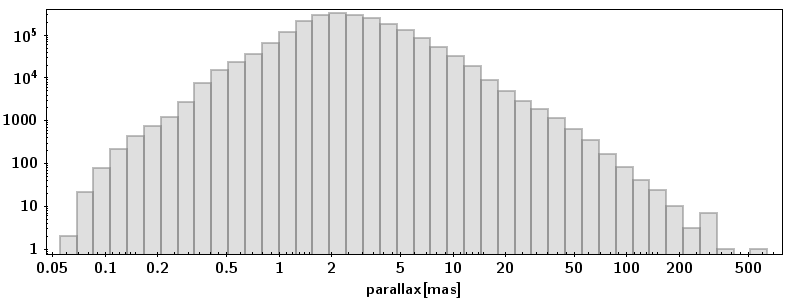}
\includegraphics[width=0.45\textwidth]{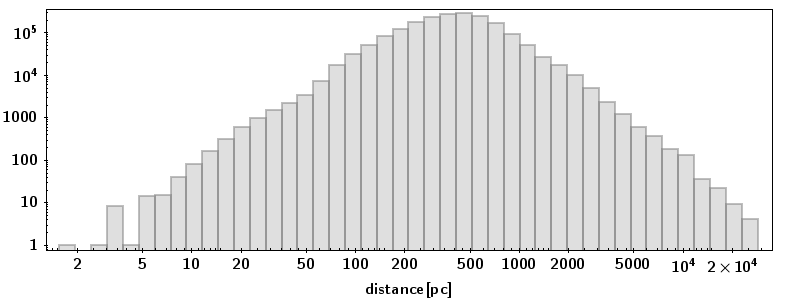}
\includegraphics[width=0.45\textwidth]{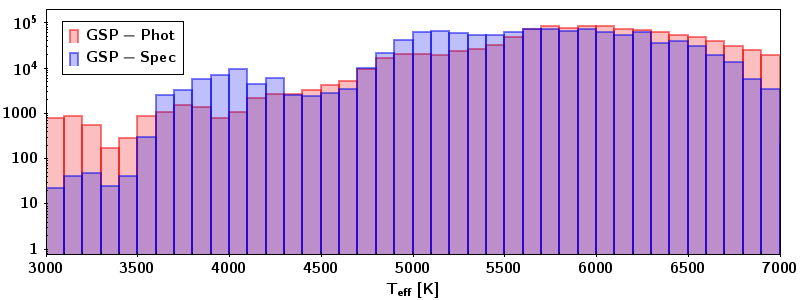}
\includegraphics[width=0.45\textwidth]{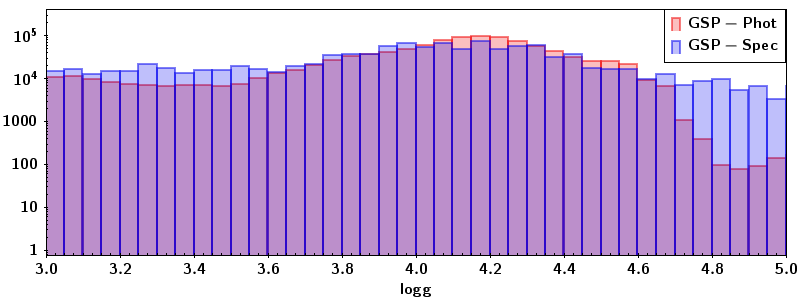}
\includegraphics[width=0.45\textwidth]{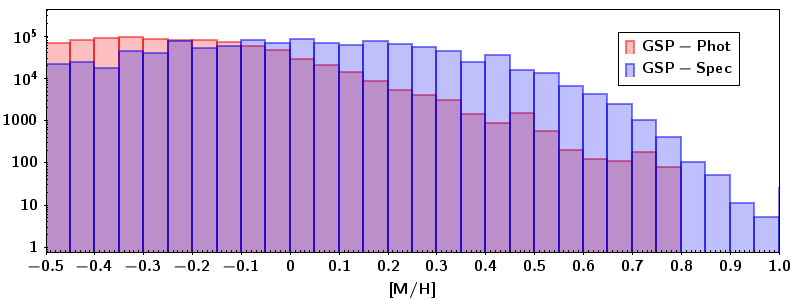}
\caption[]{Distribution in \gmag, \parallax, distance \citep[as derived by GSP-Phot][]{DR3-DPACP-156}, \teff, \logg, and \mh\ for the sources for which the activity index has been derived.}
\label{fig:cu8par_apsis_espcs_gen_distr}
\end{center}
\end{figure*}

The Apsis \espcs\ module takes the RVS average continuum normalised spectra in the stellar rest-frame as input \citep[see][for details on the Apsis input data]{onlinedocdr3,DR3-DPACP-157}.
In order to set up a template spectrum representative of the inactive photosphere, knowledge of the stellar fundamental parameters (\teff, \logg, \mh) is required (see Sect.\,\ref{sec:templates}). 
These parameters are taken from the output of either the General Stellar Parametrizer from Spectroscopy \gspspec\ module \citep{DR3-DPACP-186} or the General Stellar Parametrizer from Photometry \gspphot\ module \citep{DR3-DPACP-156,2012MNRAS.426.2463L,2011MNRAS.411..435B}).
During the processing, the default value is to use the parameters from \gspspec\ because the activity index is derived from the same data, but when they are not available, the ones from \gspphot\ are used.
The projected rotational velocity, {\fieldName{vbroad}}, provided by CU6 is also used when available (see Sect.\,\ref{sec:templates}).
Filtering of the input data is described in \cite{onlinedocdr3}.

In Fig.\,\ref{fig:cu8par_apsis_espcs_gen_distr} we show the distribution in apparent \gmag\ magnitude, parallax (\parallax), distance \citep[as derived by GSP-Phot][]{DR3-DPACP-156}, \teff, \logg, and \mh\ for the sources for which the activity index has been derived.

\subsection{Reference inactive spectrum}\label{sec:templates}

We adopt a purely photospheric theoretical spectrum as reference inactive spectrum.
The motivations and limitations of this choice are discussed extensively in Appendix\,\ref{sec:activity_indices}.
Here we outline that this is a practical and robust assumption that allows us to put stars on a relative and homogeneous activity scale.

\begin{figure*}
\begin{center}
\includegraphics[width=0.45\textwidth]{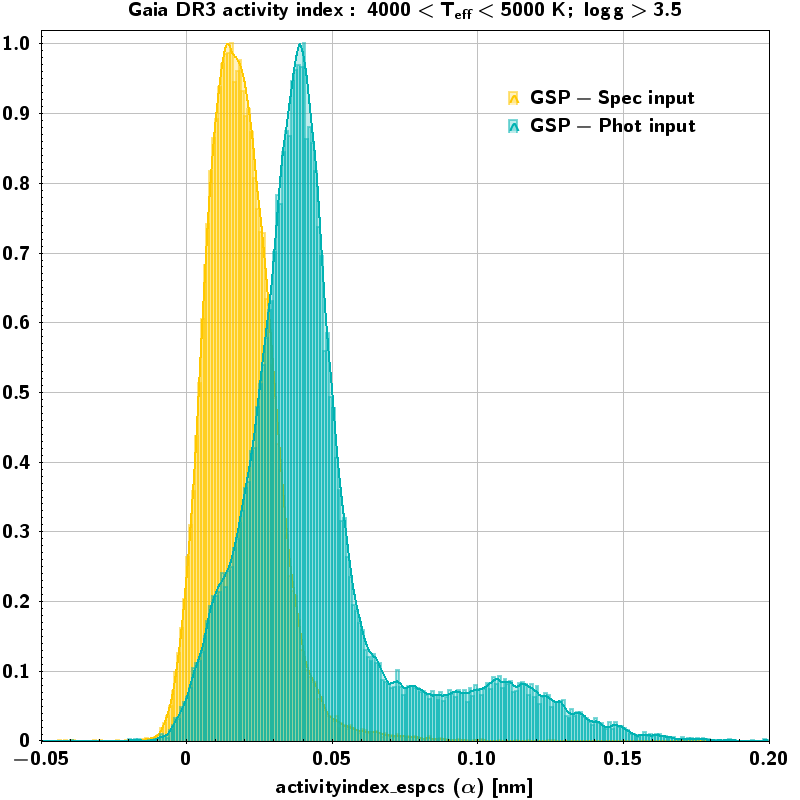}
\includegraphics[width=0.45\textwidth]{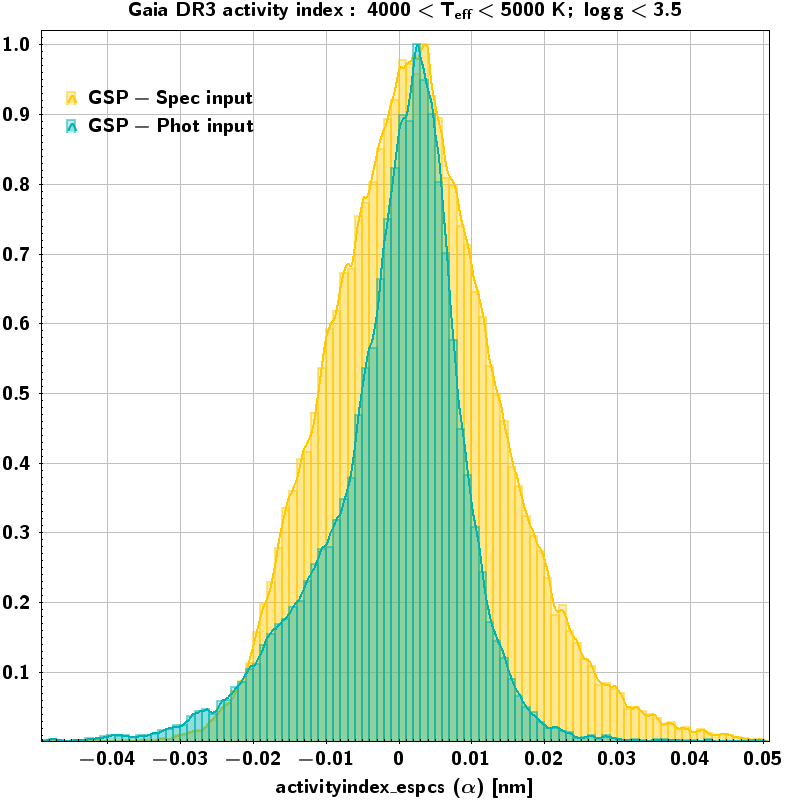}
\caption[]{Activity index histogram (bin size 0.001 \nm, with KDE superimposed) normalised to peak values of stars with $T_{\rm eff} \in (4000\,K, 5000\,K)$. The left panel includes  stars with $\log g > 4.0$ (dwarfs) and the right panel stars with $\log{g} \in (3.0, 3.5)$. From this latter we estimate an upper limit to the true dispersion of the sample of 0.011\,\nm and 0.006\,\nm for the \gspspec\ and \gspphot\ input respectively, and an upper limit to the true bias of $\simeq$ 0.002\,\nm (see text for details).}
\label{fig:cu8par_apsis_espcs_hyst}
\end{center}
\end{figure*}

The purely photospheric theoretical spectrum is obtained by a linear interpolation over a grid of MARCS synthetic spectra at the stellar parameters (\teff, \logg, \mh) as determined by \gspspec\ or \gspphot\ (Sect.\,\ref{sec:data}).
Linear interpolation is preferred to more elaborate spectrum representations because of the low computational cost combined with a reproduction of the observed spectrum which is satisfactory for our purposes (see, e.g., Figs.\,\ref{fig:cu8par_apsis_espcs_hyst} and \ref{fig:sample_spectra}).
The synthetic spectra are computed under the assumption of local thermodynamic equilibrium (LTE) and radiative equilibrium (RE).
As discussed in Appendix\,\ref{sec:activity_indices}, deviations from the LTE approximation are expected to be small in the range of parameters explored and not critical for our purposes.
Also the assumption of RE does not pose critical issues in putting stars on a relative activity scale, despite the fact that the upper photosphere is not expected to be strictly in radiative equilibrium even in the most inactive stars.

After interpolation, the reference spectrum is rotationally broadened at the projected rotational velocity \vsini\ derived by the CU6 analysis (parameter {\fieldName{vbroad}}). 

The interpolated rotationally broadened synthetic spectrum is taken as reference (or template) inactive spectrum. 

\subsection{Activity index}
\label{sec:activity_index}

The activity index is derived by comparing the observed RVS spectrum with the radiative equilibrium model spectrum computed as described in the previous section.

An excess equivalent width factor in the core of the \ion{Ca}{ii}\,IRT lines, computed on the observed-to-template ratio spectrum, is taken as an index of the stellar chromospheric activity or, in more extreme cases, of the mass accretion rate in pre-main sequence stars (see Appendix\,\ref{sec:mass_accretion}).

Given the template spectrum, the observed spectrum is first re-normalised by multiplying it by a third order Chebyshev polynomial, whose coefficients are determined by minimising the pixel-to-pixel average deviation. 
Then the observed-to-template ratio spectrum is calculated on a common wavelength grid at the stellar rest frame. 
The ratio spectrum is integrated on a $\pm \Delta \lambda = 0.15 \mathrm{nm}$ interval around the core of each of the triplet lines and the mean of the 3 values obtained is taken as activity index.
Formally, the activity index ($\alpha$, parameter {\fieldName{activityindex_espcs}} in \nm units in the \gdr{3} catalog) is defined as
\begin{equation}
\label{eq:cu8_espcs_activityIndex_def}
   \alpha \equiv \bigintsss_{-\Delta \lambda}^{+\Delta \lambda}  \left( \frac{r_{\lambda}^{\mathrm{obs}} }{ r_{\lambda}^{\mathrm{templ} }} - 1 \right) d \lambda \, ,
\end{equation}
where $r_{\lambda}^{\mathrm{obs}}$ and  $r_{\lambda}^{\mathrm{templ}}$ are the observed and the template spectrum, respectively, normalised to the continuum,
\begin{equation}
r_{\lambda} \equiv \frac{f_{\lambda}}{f_{\lambda}^c} \,.
\end{equation}
Uncertainties (parameter {\fieldName{activityindex_espcs_uncertainty}} in \nm units in the \gdr{3} catalog) are evaluated by propagating the pixel standard error, ignoring theoretical uncertainties.

The parameter $\alpha$ is close to zero for inactive stars and positive for active stars, values above $\simeq$ 0.03--0.05 \nm indicating a possible contribution to the line flux due to mass accretion processes (Sect.\,\ref{sec:validation_accretion}) or enhanced activity in close binaries due to tidal interaction.
Small negative values may derive from uncertainties in the input parameters or be physically plausible for low level of chromospheric activity (see Appendix\,\ref{sec:activity_indices}). 

The activity index $\alpha$ defined in Eq.\,\eqref{eq:cu8_espcs_activityIndex_def} is suitable for comparing activity in stars with similar fundamental parameters (\teff, logg, \mh), but unsuitable for the comparison in stars with significantly different fundamental parameters, especially those with different \teff. 
In fact, a given value of $\alpha$ corresponds to significantly different values of the chromospheric flux contribution 
when the underlying continuum flux is significantly different.
In order to overcome the difficulties posed by this ``contrast'' effect, an index $R'_{\rm IRT}$ conceptually similar to \rprimeHK\ can be derived from $\alpha$ and the stellar parameters.
First, note that $\alpha$ is related to the excess equivalent width
\begin{equation}
\label{eq:cu8_espcs_excess_ew}
   \Delta W \equiv \int_{-\Delta \lambda}^{+\Delta \lambda}  \left( r_{\lambda}^{\mathrm{obs}} - r_{\lambda}^{\mathrm{templ}} \right) d \lambda \, ,
\end{equation}
by the relation:
\begin{equation}
\label{eq:cu8_espcs_excess_ew_alpha}
    \Delta W \simeq \left\langle r_{\lambda} \right\rangle_{\mathrm{core}} \alpha
\end{equation}
where $\langle r_{\lambda} \rangle_{\mathrm{core}}$ is the expected line-to-continuum average flux ratio in the core ($\pm 0.15$ nm from central $\lambda$) of the \ion{Ca}{ii}\,IRT lines in radiative equilibrium.
The activity flux ${\cal F}'$ at the stellar surface, being either chromospheric activity or accretion flux, can be obtained by the relation:
\begin{equation}
\label{eq:cu8_espcs_activity_flux}
    {\cal F}' = {\cal F}_c \Delta W 
              \simeq {\cal F}_c \left\langle r_{\lambda} \right\rangle_{\mathrm{core}} \alpha
\end{equation}
where ${\cal F}_c$ is the continuum flux at the stellar surface\footnote{Note that by expressing ${\cal F}_c$ in [W m$^{-2}$ nm$^{-1}$], ${\cal F}'$ has units [W m$^{-2}$], expressing the chromospheric contribution to the flux in the line integrated over a wavelength range around the core of the line.}.
By analogy to \rprimeHK, the $R'_{\rm IRT}$ index can therefore be defined as
\begin{equation}
\label{eq:rp_irt_definition}
R'_{\rm{IRT}} \equiv \frac{\mathcal{F'}}{\sigma T_{\rm eff}^4} \simeq \left( \frac{{\cal F}_c \left\langle r_{\lambda} \right\rangle_{\mathrm{core}}}{\sigma T_{\rm eff}^4} \right) \alpha
\end{equation}
which gives the flux difference with respect to the radiative equilibrium approximation in the core of the \ion{Ca}{ii}\,IRT lines in bolometric flux units and, for $\alpha>0$, is suitable for comparing the activity level over the whole range of parameters explored (see Appendix\,\ref{sec:activity_indices}).

The quantity in brackets in Eq.\,\eqref{eq:rp_irt_definition} depends mainly on \teff, it is weakly dependent on metallicity and has a negligible dependence on \logg\ in the range of parameters explored in \gdr{3}.
We approximated the $\log$ of this quantity with a third order polynomial in $\theta = \log{T_{\rm eff}}$ for four values of \mh.
The fit coefficients are reported in Table\,\ref{tab:poly_coefficients}.
Using this polynomial fit, $R'_{\rm{IRT}}$ can be estimated from $\alpha$ and \teff\ for a given \mh\ as:
\begin{equation}
\label{eq:conversion}
\log R'_{\rm{IRT}} \simeq (C_0 + C_1 \theta + C_2 \theta^2 + C_3 \theta^3) + \log \alpha
\end{equation}

\begin{table}
   \centering
      \caption{Polynomial coefficients for converting $\alpha$ to $\log R'_{\rm{IRT}}$ using Eq.\,\eqref{eq:conversion}.}
      \label{tab:poly_coefficients}
      \begin{tabular}{@{}rllll@{}}
      \hline
\mh     &  $C_0$ & $C_1$ & $C_2$ & $C_3$ \\
\hline
-0.50    & -3.3391 &   -0.1564 &  -0.1046  &  0.0311 \\  
0.00     & -3.3467 &   -0.1989 &  -0.1020  &  0.0349 \\
0.25     & -3.3501 &   -0.2137 &  -0.1029  &  0.0357 \\ 
0.50     & -3.3527 &   -0.2219 &  -0.1056  &  0.0353 \\ 
\hline
      \end{tabular}
\end{table}

\section{Scientific validation}
\label{sec:results}

\espcs\ has processed stars with $G<13$, \teff\ in the range (3000\,K, 7000\,K), \logg\ in the (3.0, 5.5) range, and \mh\ in the (-1.0, 1.0) range.
The activity index $\alpha$ has been estimated on a total of approximately 2M stars; in DR3 $\alpha$ has a range $\approx$ (-0.1, 1.0).
As an example, Fig.\,\ref{fig:cu8par_apsis_espcs_hyst} reports the activity index histogram for stars with $T_{\rm eff} \in (4000\,K, 5000\,K)$.
Negative values are physically expected because in low activity stars the chromosphere causes a deeper line absorption than the radiative equilibrium expectation (see Appendix\,\ref{sec:activity_indices}); uncertainties in the astrophysical parameters can also cause negative values. 
Nominal uncertainties take the pixel noise into account and are $\sim$ 0.001 nm on average.

The activity index for the bulk of K stars with the lowest \logg\ ($3.0 \le \log g \le 3.5$) has standard deviation of 0.011 and 0.006 nm in the case of \gspspec\ and \gspphot\ input, respectively (Fig.\,\ref{fig:cu8par_apsis_espcs_hyst}, right panel).
Since these stars, which exclude sub-giants in close active binaries like RS\,CVn systems, have the lowest activity in the parameter range explored, such standard deviations can be taken as upper limits of the random errors of the whole sample.
The average activity index for these stars is 0.002\,\nm for both the \gspphot\ and \gspspec\ input; this can be considered an upper limit to the true bias.

Astrophysical parameters for rapidly rotating and emission line stars are generally provided by \gspphot\ only.
Since these are also the most active stars, the distribution with \gspphot\ input is peaked and extended to larger values than the distribution of the activity index with \gspspec\ input.

\subsection{Activity index vs \teff}
\label{sec:activity_vs_teff}

\begin{figure*}[ht]
\begin{center}
\includegraphics[width=0.45\textwidth]{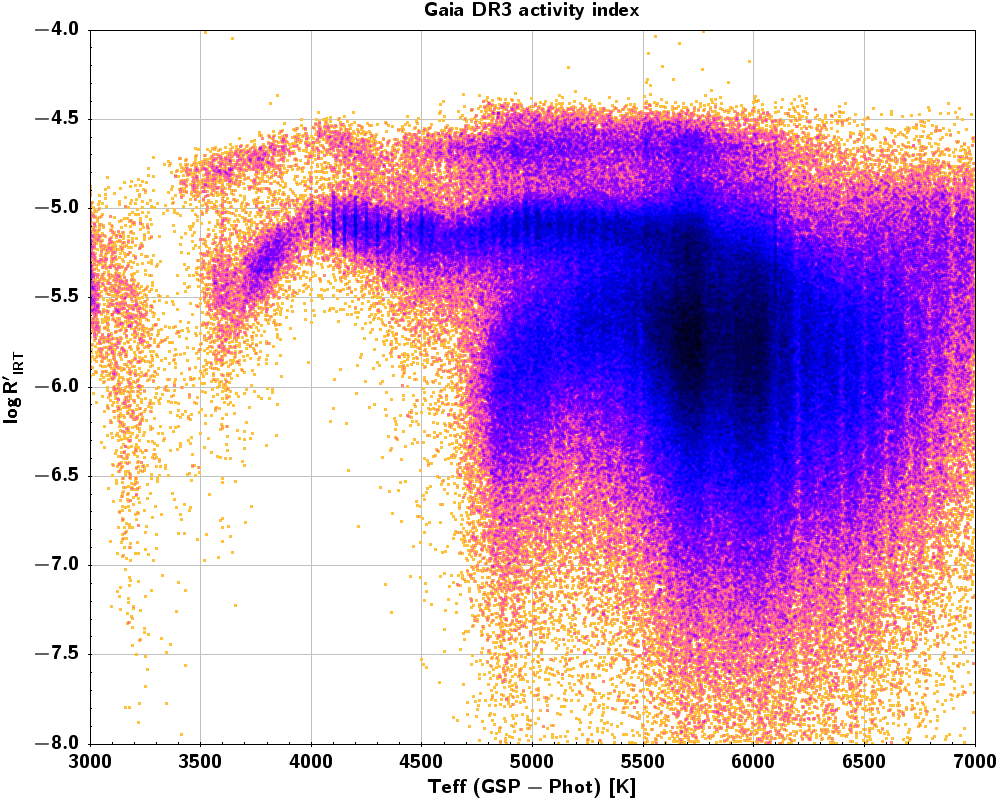}
\vspace{0.5cm}
\includegraphics[width=0.45\textwidth]{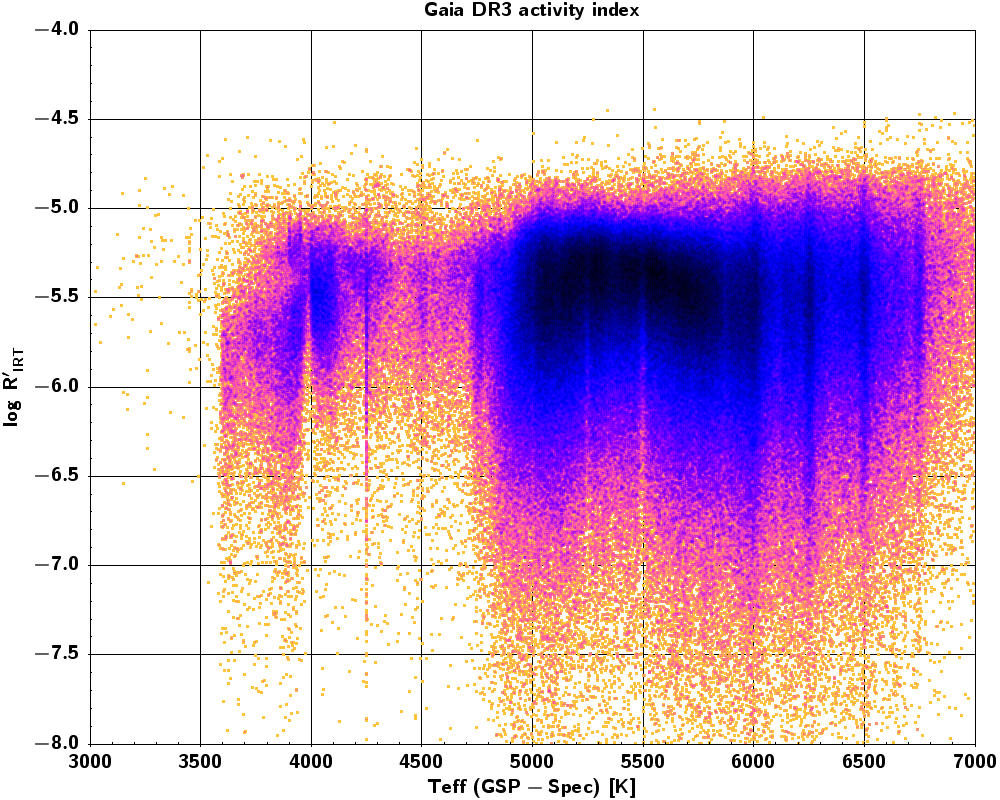}
\vspace{0.5cm}
\includegraphics[width=0.45\textwidth]{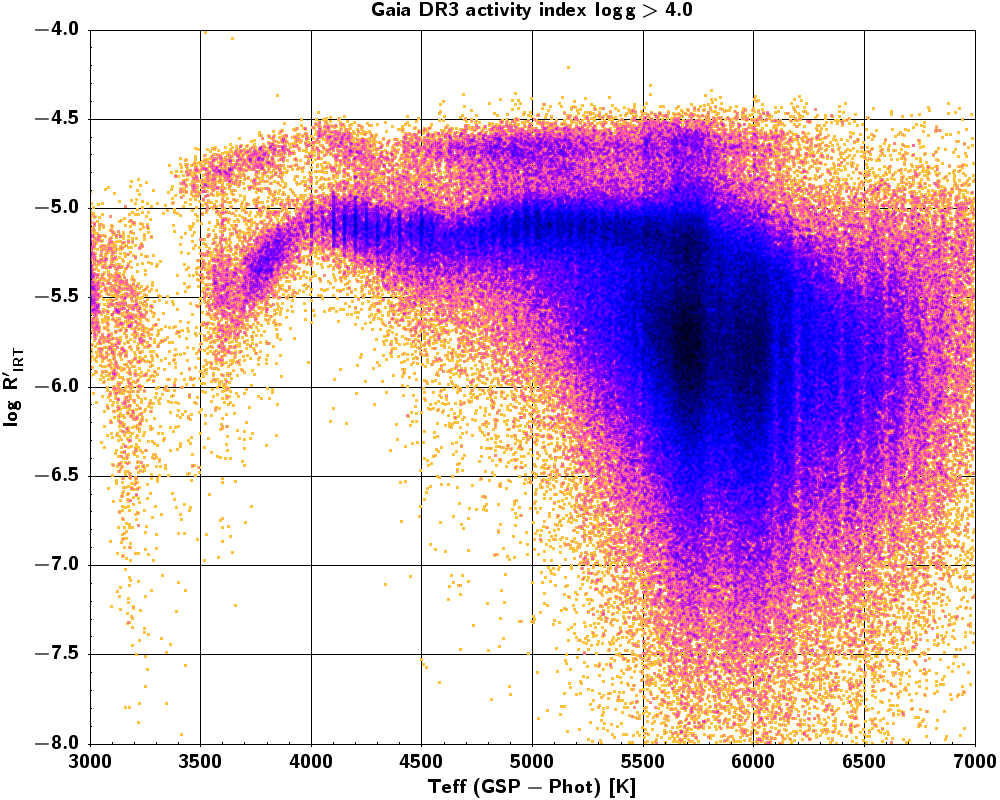}
\includegraphics[width=0.45\textwidth]{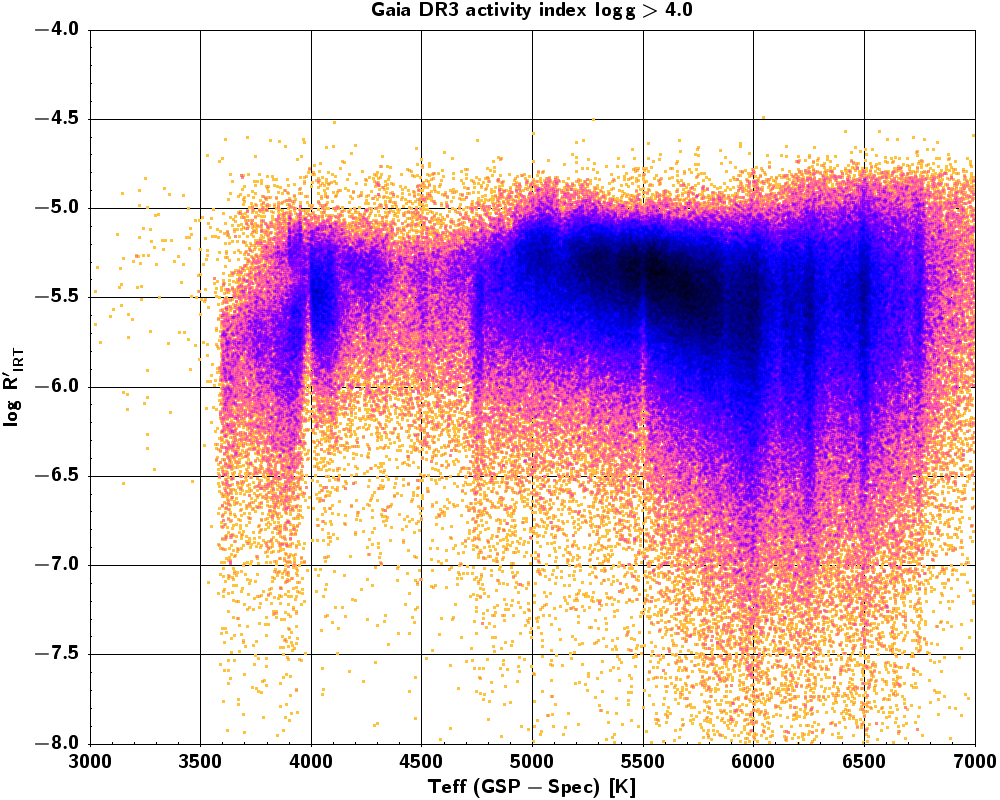}
\includegraphics[width=0.45\textwidth]{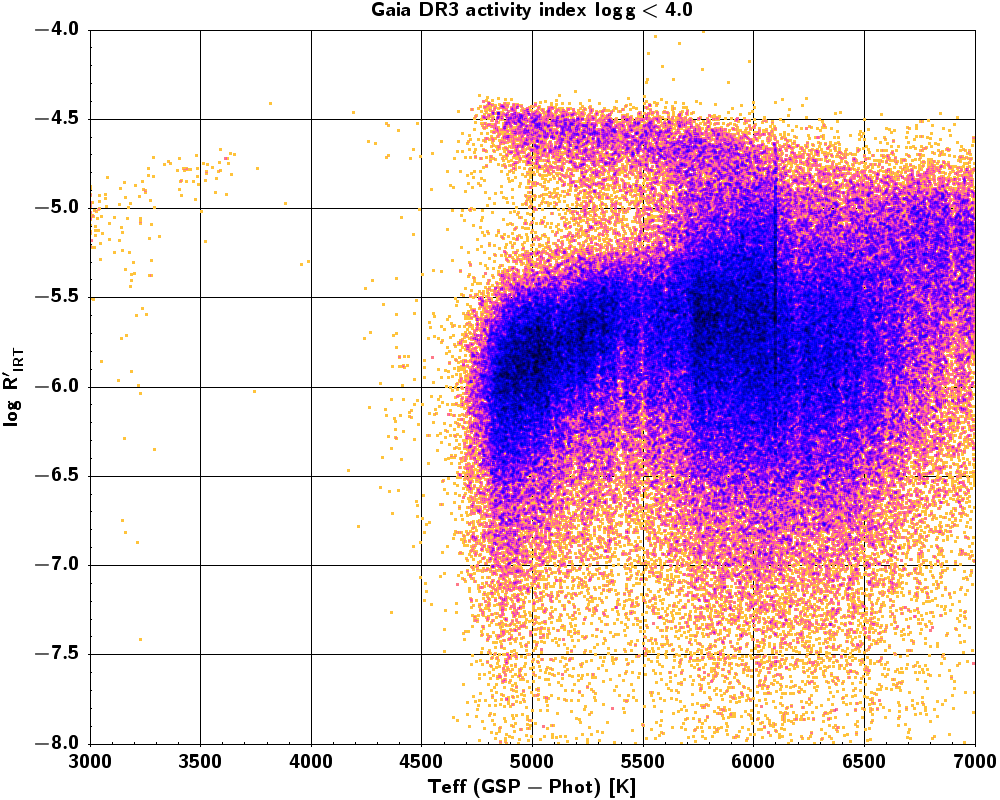}
\includegraphics[width=0.45\textwidth]{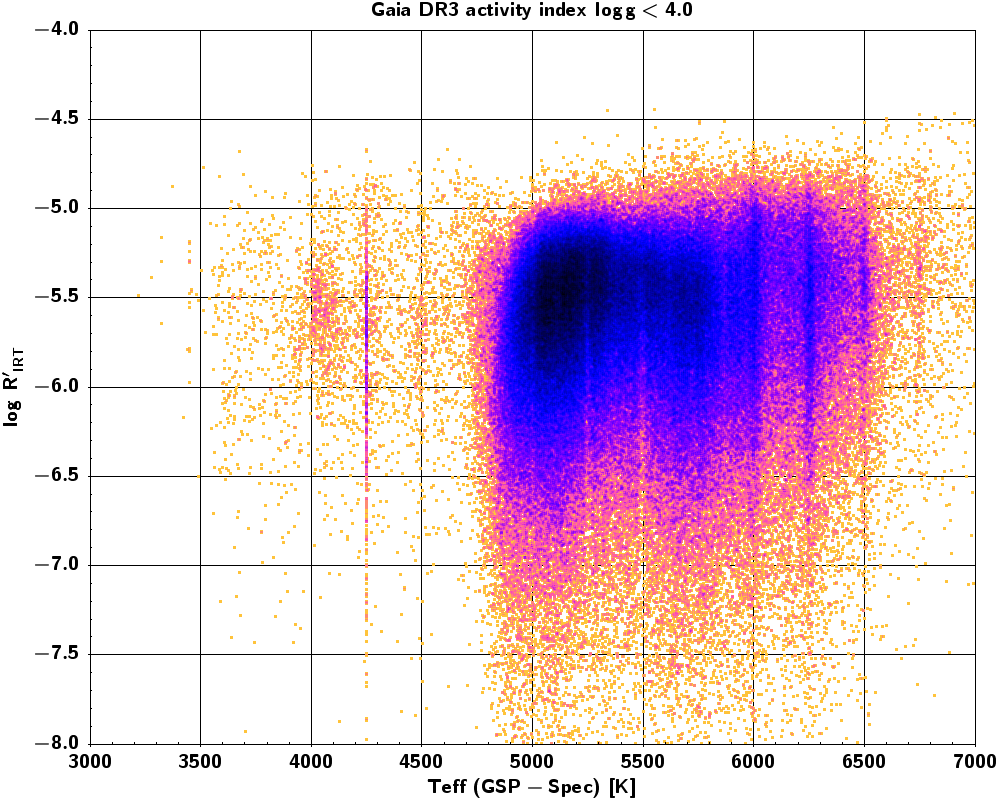}
\caption[]{Density plots of $\log R'_{\rm IRT}$ vs \teff\ with input parameters from \gspphot\ (upper left panel) and from \gspspec\ (upper right panel). The same plots are repeated in the middle panels for $\log g > 4.0$ and in the lower panels for $\log g < 4.0$.}
\label{fig:RpIRTvsTeff}
\end{center}
\end{figure*}

\begin{figure}[h]
\begin{center}
\includegraphics[width=0.45\textwidth]{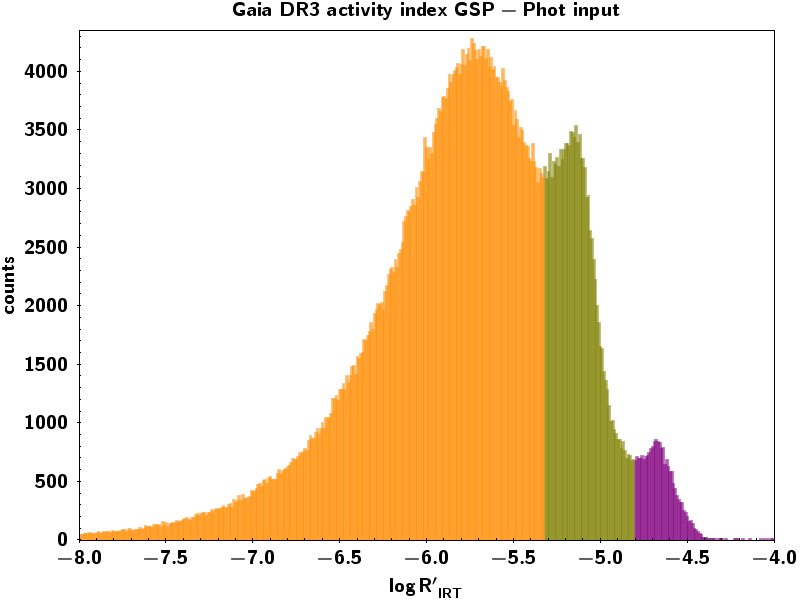}
\caption[]{Histogram of $\log{R'_{\rm IRT}}$ with input from \gspphot\ (bin size 0.01). The highest branch (VHA) is coloured in purple and has a lower limit between -4.8 and -5.0, depending on \teff. The second highest branch (HA) is coloured in green and has a lower limit between -5.3 and -5.5, depending on \teff. The remaining of the distribution (LA) is coloured in orange.}
\label{fig:RpIRT_gspphot_histo}
\end{center}
\end{figure}

Fig.\,\ref{fig:RpIRTvsTeff} shows the 2D  density plot $\log R'_{\rm IRT}$ vs \teff\ with input parameters from \gspphot\ and \gspspec\ separately.
In the dataset with input parameters from \gspphot\ (Fig.\,\ref{fig:RpIRTvsTeff}, upper left panel), the high activity part, say for $\log R'_{\rm IRT} \gtrsim -5.5$, is characterised by two almost horizontal branches, bending towards lower $R'_{\rm IRT}$ values at decreasing \teff\ below 4000\,K and disappearing below 3500\,K.

The clustering of $R'_{\rm IRT}$ in these three different regimes is further outlined in the histogram shown in Fig.\,\ref{fig:RpIRT_gspphot_histo}.
The highest branch (very high activity, VHA, hereafter) has a lower limit between $\log R'_{\rm IRT} \simeq -4.8$ and -5.0, depending on \teff. 
The second highest branch (high activity, HA, hereafter) has a lower limit between $\log R'_{\rm IRT} \simeq -5.3$ and -5.5, depending on \teff.
The limit between the second highest branch and the bulk of low-activity stars (LA hereafter) is at $\log R'_{\rm IRT} \simeq -5.35$

There is a clear gap between the HA-branch and the LA-branch starting at \teff$\simeq$ 5400\,K and $\log R'_{\rm IRT} \simeq -5.35$ and growing at decreasing \teff, until it disappears at \teff$\simeq$ 4500\,K (upper left panel of Fig.\,\ref{fig:RpIRTvsTeff}). 
This gap, however, disappears when considering stars with $\log g > 4.0$ only (Fig.\,\ref{fig:RpIRTvsTeff}, middle left panel).
The comparison between the upper panels and the middle ($\log g > 4.0$) and lower ($\log g < 4.0$) panels of Fig.\,\ref{fig:RpIRTvsTeff} reveals that the structure of low-activity stars bending down in this \teff\ range is, in fact, populated by more evolved stars, approximately in the sub-giant branch of the HRD, with a low level of chromospheric activity.  

These results are consistent with what found by \cite{1996AJ....111..439H,2021A&A...646A..77G}, who suggested a grouping of very active stars with $\log R'_{\rm HK} > -4.2$ (corresponding approximately to $\log R'_{\rm IRT} > -5.2$, according to the correlation discussed in Sect.\,\ref{sec:validation_FEROS}) and very inactive stars with $\log R'_{\rm HK} < -5.1$ (corresponding approximately to $\log R'_{\rm IRT} > -5.8$), with a regime of active stars in between these values.

The same density plot with input parameters from \gspspec\ (Fig.\,\ref{fig:RpIRTvsTeff}, upper right panel) does not show such clustering.
This is likely due to the higher dispersion of $R'_{\rm IRT}$ with input parameters from \gspspec\ (as suggested by a visual comparison between the two diagrams in Fig.\,\ref{fig:RpIRTvsTeff} and supported by the comparison with $R'_{\rm HK}$ in Sect.\,\ref{sec:validation_FEROS}) and the lack of \gspspec\ parameters for active and fast-rotating stars that populate the higher part of the diagram (see Sects.\,\ref{sec:validation_accretion} and \ref{sec:validation_rotmod}).
Note that the higher dispersion of $R'_{\rm IRT}$ with input from \gspspec\ suggested by Fig.\,\ref{fig:RpIRTvsTeff} is consistent with the higher $\alpha$ dispersion found in stars with $T_{\rm eff} \in (4000\,K, 5000\,K)$ and $\log{g} \in (3.0, 3.5)$ (see above and Fig.\,\ref{fig:cu8par_apsis_espcs_hyst}).

\begin{figure*}[ht]
\begin{center}
\includegraphics[width=0.45\textwidth]{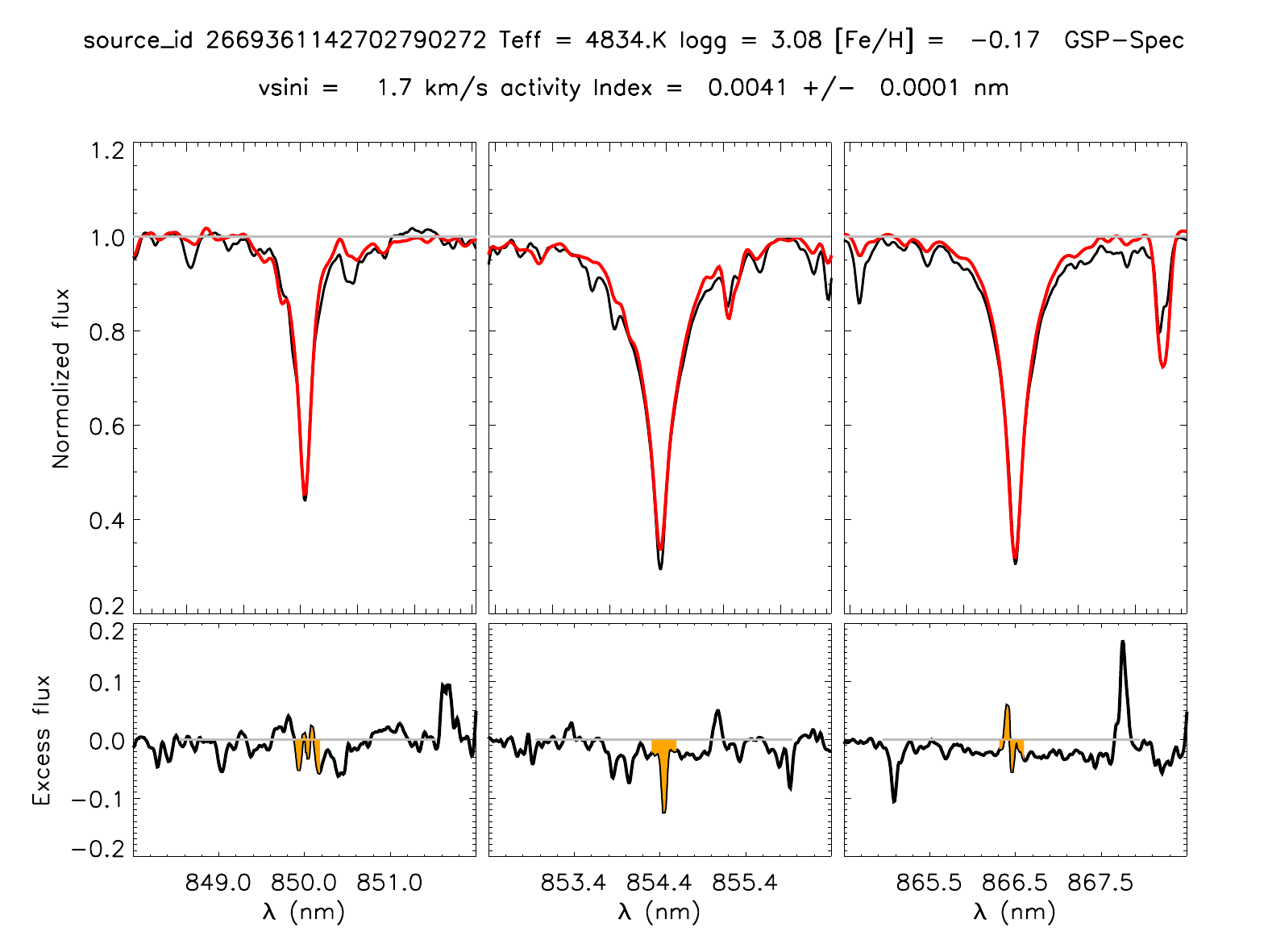}
\includegraphics[width=0.45\textwidth]{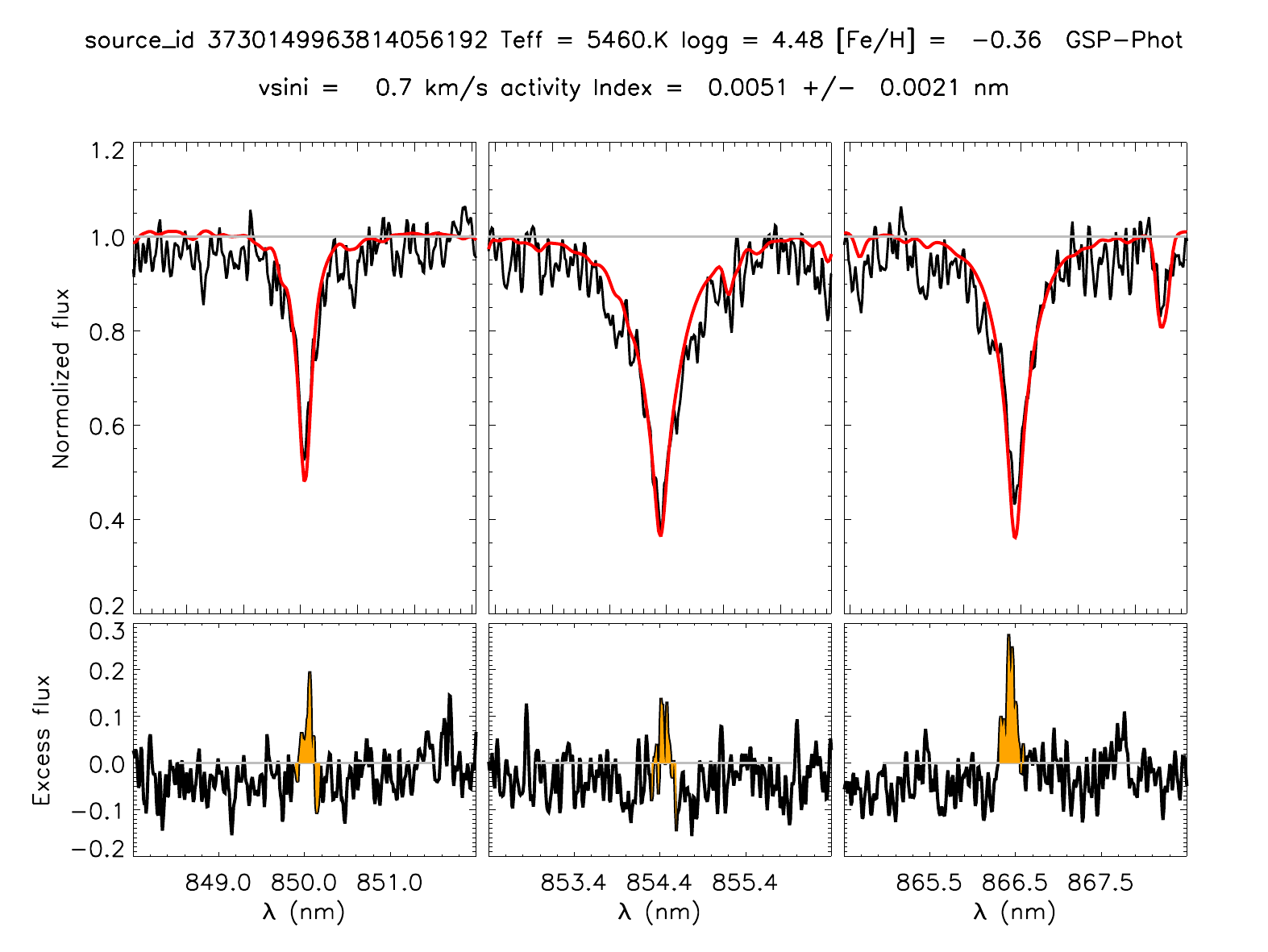}
\includegraphics[width=0.45\textwidth]{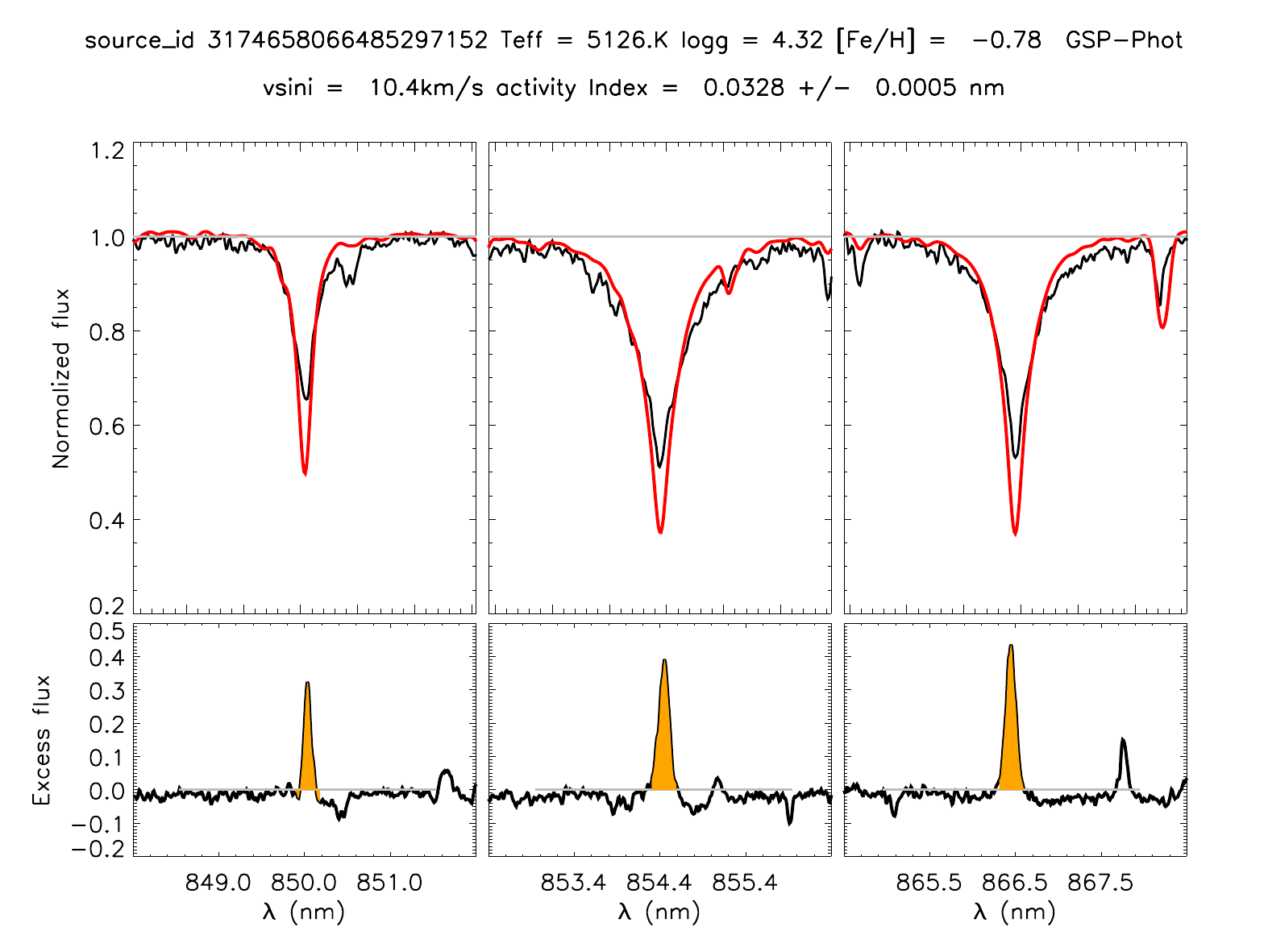}
\includegraphics[width=0.45\textwidth]{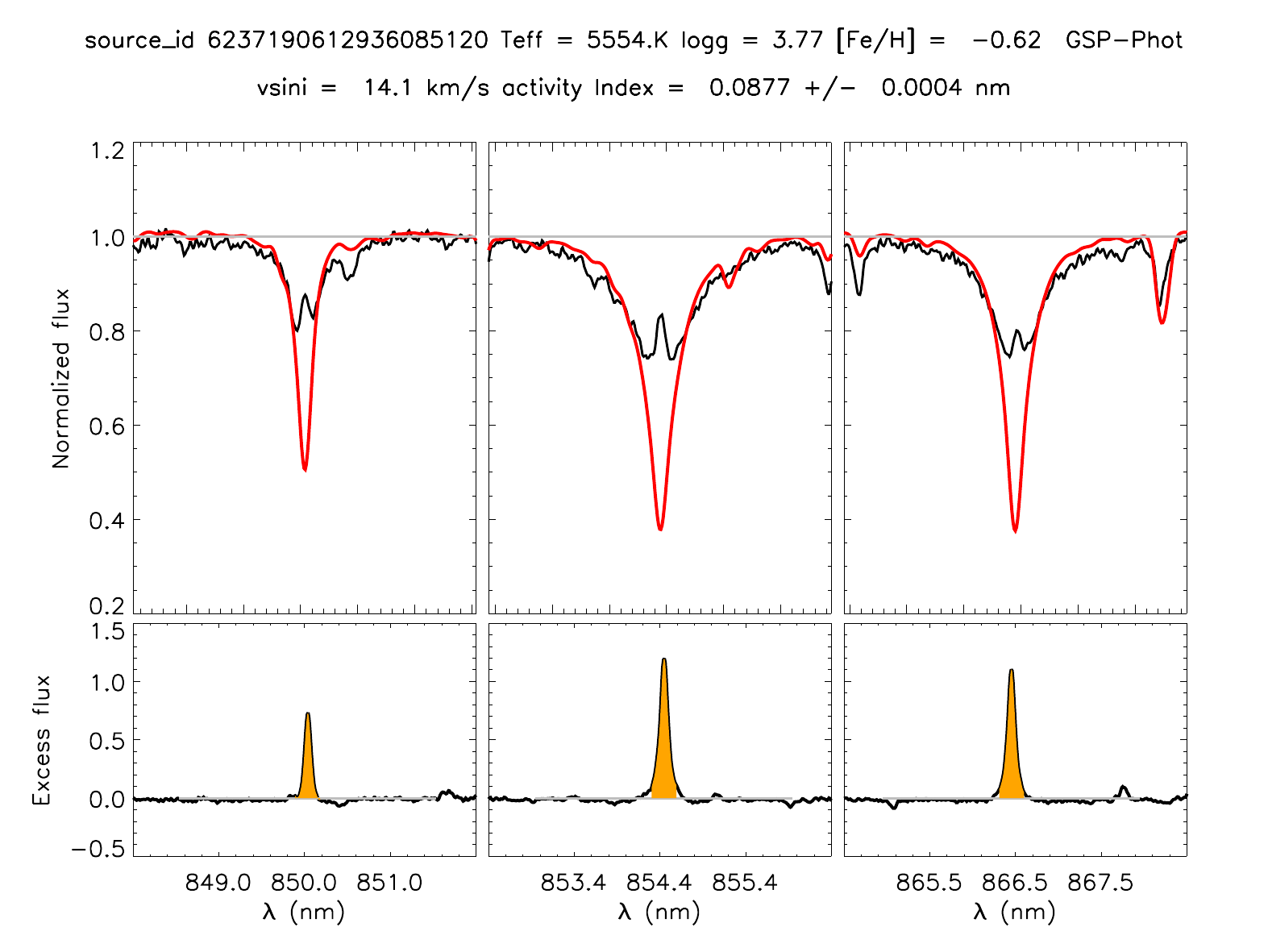}
\caption[]{\ion{Ca}{ii}\,IRT sample spectra in the different regimes. 
In each panel, the spectra (black line) are compared with the template (red line).
The enhancement factor, i.e. the integrand of  Eq.\,\eqref{eq:cu8_espcs_activityIndex_def}, is reported below the spectrum, and the integral producing $\alpha$ is outlined in orange.
Upper left panel: LA subgiant with $\log{R'_{\rm IRT}} = -6.04$.
Upper right panel: LA dwarf with $\log{R'_{\rm IRT}} = -5.98$.
Lower left panel: HA dwarf with $\log{R'_{\rm IRT}} = -5.19$
Lower right panel: an accreting star (one in the Upper Scorpius association reported in Table\,\ref{tab:accretion_ref} and Fig.\,\ref{fig:RpIRTvsMdot}) in the VHA regime.}
\label{fig:sample_spectra}
\end{center}
\end{figure*}

In Fig.\,\ref{fig:sample_spectra} some \ion{Ca}{ii}\,IRT sample spectra of stars in the different regimes are shown: a subgiant and a dwarf in the LA regime, a dwarf in the HA regime and an accreting star (one in the Upper Scorpius association reported in Table\,\ref{tab:accretion_ref} and Fig.\,\ref{fig:RpIRTvsMdot}) in the VHA regime.
The spectra are compared with the template and the enhancement with respect to the template spectrum is shown.

The activity distribution is found to be very different for \teff$\apprle$\,3500\,K, which corresponds to the transition between partially-convective to fully convective stars.
Fig.\,\ref{fig:RpIRTvsTeff} shows that stars close to the partially-convective / fully convective boundary have a large drop in chromospheric activity, which increases again with decreasing \teff\ (mass) in the fully-convective regime.
A paucity of stars is found in the \teff -$\log R'_{\rm IRT}$ diagram at 3300\,K$\apprle$\teff$\apprle$\,3500\,K, likely connected to the gap in the mid-M dwarfs main sequence found by \cite{2018ApJ...861L..11J}, linked
to the onset of full convection in M dwarfs \citep{2018ApJ...861L..11J,2010MNRAS.408.1409M,2018MNRAS.480.1711M,2018A&A...619A.177B}.

\begin{figure*}
\begin{center}
\includegraphics[width=0.48\textwidth]{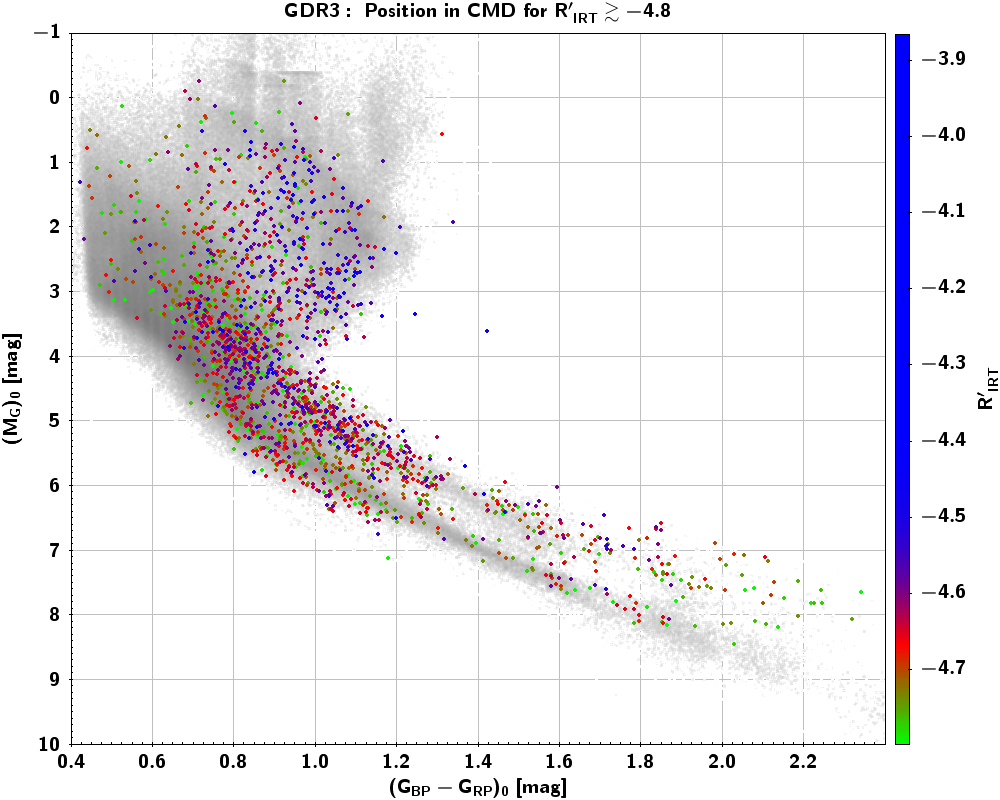}
\includegraphics[width=0.48\textwidth]{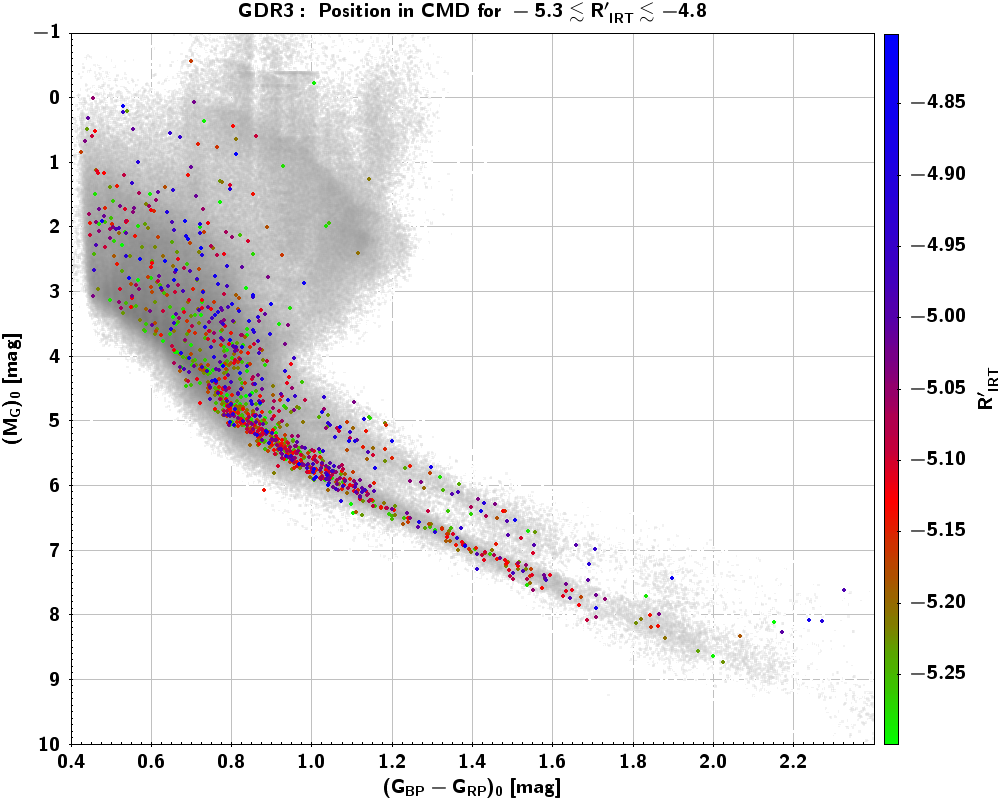}
\includegraphics[width=0.48\textwidth]{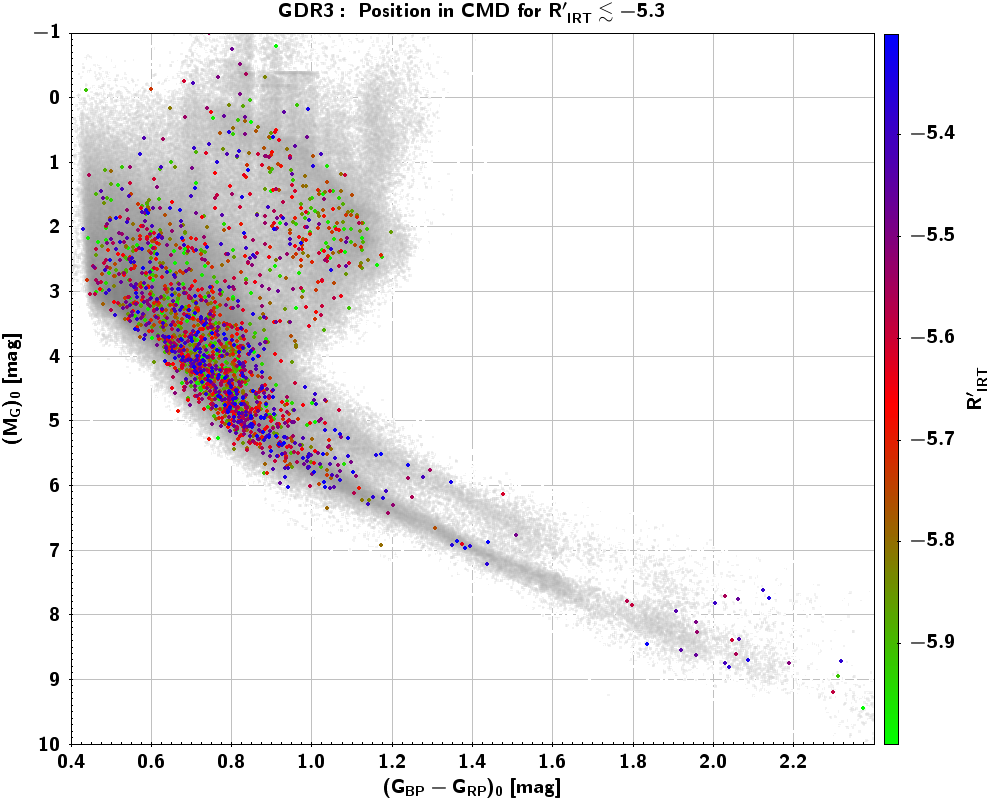}
\caption[]{Position in the CMD, corrected for interstellar extinction, of stars in the VHA (upper left panel), HA (upper right panel) and LA (lower panel) branches. Input APs are from \gspphot.
The density plot of all sources for which the activity index has been derived is shown in grey on the background.
The position of sub-samples of $\sim 2000$ stars per branch are shown in each panel, colour coded according to $R'_{\rm IRT}$.
See text for details.}
\label{fig:ai_CMD_gspphot}
\end{center}
\end{figure*}

\subsection{Comparison with the $R'_{\rm HK}$ chromospheric activity index}
\label{sec:validation_FEROS}

\begin{figure}[ht]
\begin{center}
\includegraphics[width=0.45\textwidth]{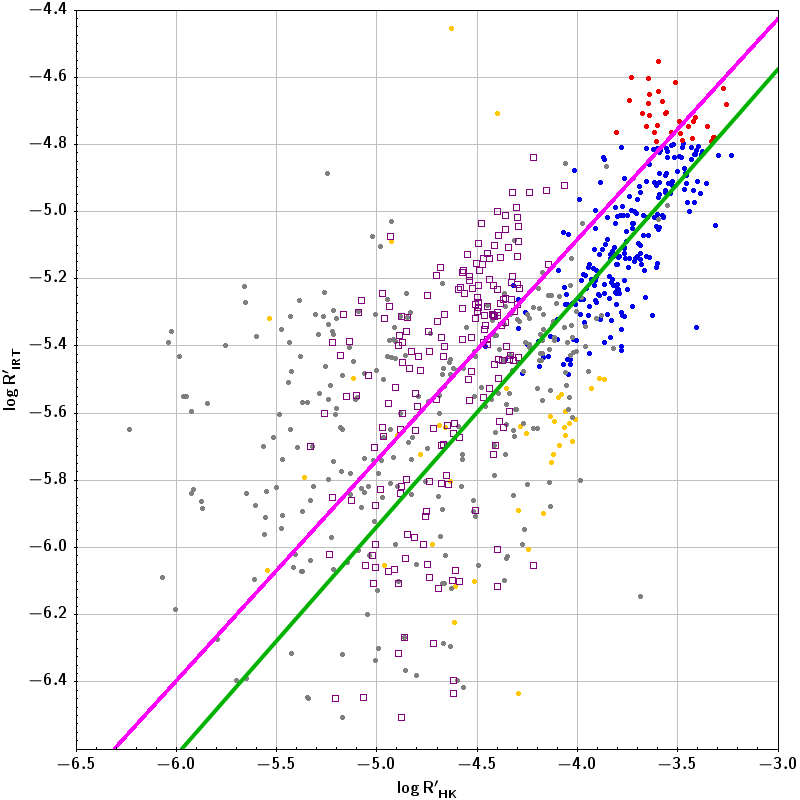}
\caption[]{Comparison of $\log R'_{\rm IRT}$ vs $\log R'_{\rm HK}$ obtained from ESO-FEROS archive spectra and with $\log R'_{\rm HK}$ from \cite{2018A&A...616A.108B}. Stars with $R'_{\rm IRT}$ obtained from \gspphot\ input and placed in the HA-branch are plotted in blue, those in the VHA-branch in red, and those in the LA-branch in orange.  Stars with $R'_{\rm IRT}$ obtained from \gspspec\ input are plotted in grey. The linear fit to data with \gspphot\ input is plotted in green (Eq.\,\eqref{eq:RpIRTvsRpHKPhot}). The comparison with  $R'_{\rm HK}$ from \cite{2018A&A...616A.108B} is shown with open purple squares, and the corresponding linear fit with a purple line (Eq.\,\eqref{eq:RpIRTvsRpHKBS}).}
\label{fig:RpIRTvsRHK}
\end{center}
\end{figure}

In order to relate $R'_{\rm IRT}$ with $R'_{\rm HK}$, we have estimated $R'_{\rm HK}$ on ESO-FEROS archive spectra and further considered the data provided by \cite{2018A&A...616A.108B}, which includes a compilation of $R'_{\rm HK}$ values from various surveys plus their own measurements on archive HARPS spectra.
For simplicity, our approximate estimate of $R'_{\rm HK}$ from FEROS spectra have been carried out as in  \cite{1979ApJS...41...47L}, without considering more recent calibrations of the H \& K surface flux \citep[][]{1982A&A...113....1M,1984A&A...130..353R} nor rescaling to the Mount Wilson data as in \cite{2018A&A...616A.108B}. 

Deriving an accurate $R'_{\rm IRT}$ vs $R'_{\rm HK}$ relationship would require a consistent estimate of the photospheric contribution and a homogenisation of the methods used in deriving $R'_{\rm HK}$, considering also the non-simultaneity of the measurements and the intrinsic variability of chromospheric activity.
This is outside the scope of the present work and is left for future work, together with the publication of the $R'_{\rm HK}$ values from archive FEROS spectra.
Here we just aim at verifying the existence of the correlation between $R'_{\rm IRT}$ and $R'_{\rm HK}$ and at providing some preliminary approximate relationships between the two parameters.

Cross-match with the \gdr{3} activity index leads to 671 objects in common with our $R'_{\rm HK}$ estimate from FEROS spectra (362 with \gspspec\ and 309 with \gspphot\ input parameters) and 220 objects in common with the \cite{2018A&A...616A.108B} dataset.
As discussed above (see Figs.\,\ref{fig:cu8par_apsis_espcs_hyst} and \ref{fig:RpIRT_gspphot_histo}), $R'_{\rm IRT}$ with \gspspec\ input is characterised by a larger dispersion and is more concentrated on the lower activity range, while $R'_{\rm IRT}$ with \gspphot\ input has a lower dispersion and extends to a higher activity range.
The cross-match with the FEROS data reflects this situation; in Fig.\,\ref{fig:RpIRTvsRHK}, showing the $R'_{\rm IRT}$ vs $R'_{\rm HK}$ dispersion diagram, data with \gspspec\ input covers mostly the lower left part while data with \gspphot\ input covers mostly the upper right part of the diagram.
Data with \gspspec\ input is also affected by a larger dispersion than data with \gspphot\ input or the \cite{2018A&A...616A.108B} dataset.

The $R'_{\rm IRT}$ with \gspphot\ input is well correlated with FEROS $R'_{\rm HK}$, with a Pearson linear correlation coefficient $r=0.73$ and
\begin{equation}
\label{eq:RpIRTvsRpHKPhot}
\log R'_{\rm IRT} = 0.68066 \log R'_{\rm HK} - 2.53461
\end{equation} 

The $R'_{\rm IRT}$ with \gspspec\ input is poorly correlated with FEROS $R'_{\rm HK}$, having $r=0.31$; this is mainly due to the the fact that the \gspspec\ APs in \gdr{3} concern mostly stars with a low level of activity and therefore a regime in which the activity index signal-to-noise ratio is lower.
The larger variance of $\alpha$  with \gspspec\ input with respect to the \gspphot\ input (see above and Fig.\,\ref{fig:cu8par_apsis_espcs_hyst}, right panel), could itself be due to the lower activity index signal-to-noise ratio in the regime covered by \gspspec, or to an intrinsically larger variance of the \gspspec\ APs.
Despite such limitations, the activity index derived with \gspspec\ input APs maintains its validity of a relative chromospheric activity indicator in the LA regime.

Correlation with the \cite{2018A&A...616A.108B} dataset has $r=0.50$ and
\begin{equation}
\label{eq:RpIRTvsRpHKBS}
\log R'_{\rm IRT} = 0.65691 \log R'_{\rm HK} - 2.45543
\end{equation}

\subsection{Position in the colour-magnitude diagram}\label{sec:validation_cmd}

Consistency of the activity index scale is verified also by examining the position of the three branches in the \gaia\ CMD \citep[\mg\ vs \bmr\ corrected for interstellar extinction,][]{DR3-DPACP-156}. 
Fig.\,\ref{fig:ai_CMD_gspphot} shows the position of sub-samples of each branch superimposed on the CMD of all sources for which the activity index has been derived.
The range of input APs considered implies that the CMD in Fig.\,\ref{fig:ai_CMD_gspphot} includes the lower MS, the sub-giant branch and the PMS regions.
The MS, in turn, is split in two sequences, the over-density associated with the higher one notoriously due mainly to binary stars, but with PMS stars transiting over the same region while they contract towards the ZAMS.
In the following, we shall refer to the single star MS as single-MS and to the binary star MS as binary-MS. 
In order to consider homogeneous results that include PMS and fast-rotating stars, only results with \gspphot\ input are shown.

The first panel of Fig.\,\ref{fig:ai_CMD_gspphot} shows that stars in the VHA-branch ($R'_{\rm IRT} \gtrsim -4.8$) are preferentially located above the single-MS, the higher $R'_{\rm IRT}$ the higher the tendency to be located at larger distance above the single-MS. Note, however, that some of the most active stars are located on or close to the single- and binary-star MS.
Furthermore, some of the most active stars are early-K stars at $\sim$ 2--4 mag above the MS.
The second panel of Fig.\,\ref{fig:ai_CMD_gspphot} shows that stars in the HA-branch ($-5.3 \lesssim R'_{\rm IRT} \lesssim -4.8$) are mostly concentrated on the single-MS, with some star on the binary-MS and a few above the MS.
Contrary to the VHA stars location, the region of early-K stars at $\sim$ 2--4 mag above the MS, where we can find RS\,CVn's and PMS, is essentially void of HA stars.
The third panel shows that stars in the LA-branch  ($ R'_{\rm IRT} \lesssim -5.3$) are mostly concentrated in the F- G-type MS and a few in the sub-giant region.
Consistently with the discussion in Sect\,\ref{sec:activity_vs_teff} (Fig.\,\ref{fig:RpIRTvsTeff}), the late-K, M stars region is almost void of LA stars. 
LA stars populate the early-K region at $\sim$ 2--4 mag above the MS, with some of the least active stars concentrated here.

In support of the scientific validity of the results published in \gdr{3}, in the following sections we discuss the consistency with current knowledge of the distribution of the three branches in the CMD.

\subsection{The very high activity branch}\label{sec:validation_accretion}

\begin{figure}
\begin{center}
\includegraphics[width=0.45\textwidth]{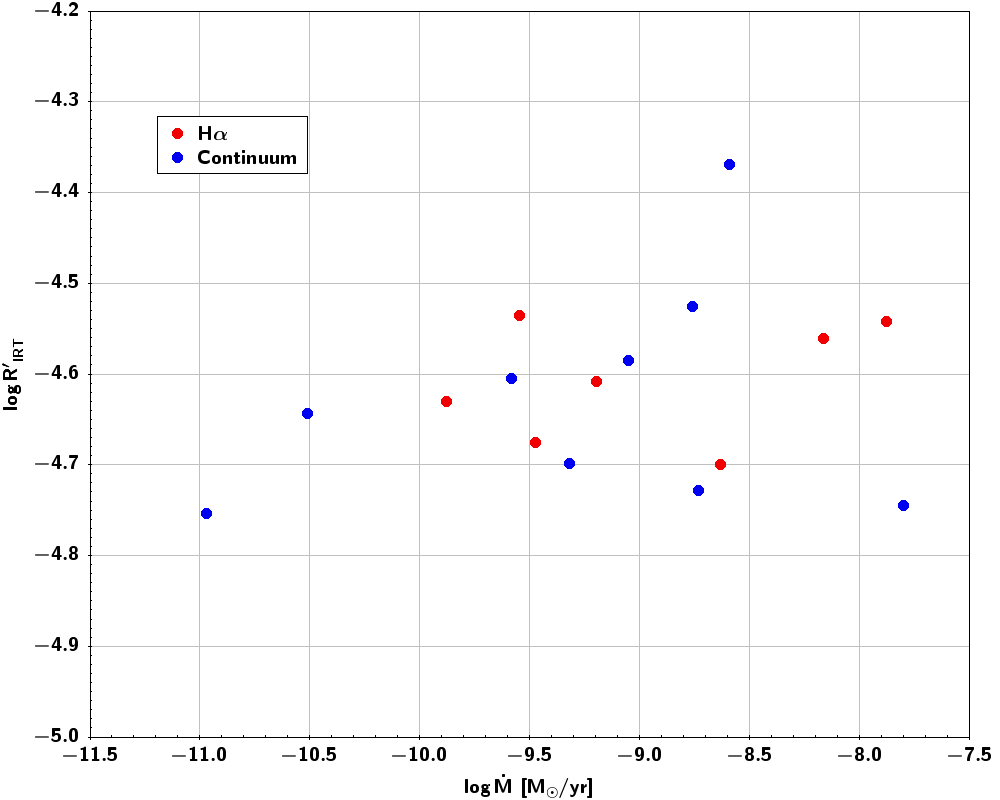}
\caption[]{Activity index ($\log R'_{\rm IRT}$) vs accretion rate ($\log{\dot{M}}$) for selected sources. Red filled circles indicate accretion rates derived from the width of H$\alpha$ at 10\% of the peak by the Gaia-ESO survey \cite[][and references therein]{2015A&A...576A..80L}. Blue filled circles indicate accretion rates derived from the excess continuum emission  \cite[see, e.g.][]{2015A&A...579A..66M}. See Table\,\ref{tab:accretion_ref} for the sources of the accretion rate estimates.}
\label{fig:RpIRTvsMdot}
\end{center}
\end{figure}

\begin{table}
\centering
\caption{Accretion rates ($\log{\dot{M}}$ in $[M_{\odot}/yr]$) and references for selected sources in Fig.\,\ref{fig:RpIRTvsMdot}. References: [1] \cite{2015A&A...576A..80L}, [2] \cite{2014A&A...561A...2A}, [3] \cite{2020A&A...639A..58M}, [4] \cite{2019A&A...632A..46V}.}
\label{tab:accretion_ref}
\begin{scriptsize}
\begin{tabular}{@{}|r|r|r|l|c|@{}}
\hline
  \multicolumn{1}{|c|}{GDR3 source\_id} &
  \multicolumn{1}{c|}{$R'_{\rm IRT}$} &
  \multicolumn{1}{c|}{$\log{\dot{M}}$} &
  \multicolumn{1}{c|}{Cluster / Association} &
  \multicolumn{1}{c|}{Ref.} \\
\hline
  5519263533510676224 & -4.63 & -9.88  & $\gamma$\,Vel.  & [1]\\
  3104345255664954368 & -4.61 & -9.19  & NGC\,2232       & [1]\\
  5201153035509861888 & -4.67 & -9.47  & Chamaeleon I    & [1]\\
  5201226389256838528 & -4.70 & -8.63  & Chamaeleon I    & [1]\\
  3340972772281842944 & -4.54 & -9.54  & $\lambda$\,Ori  & [1]\\
  3340894500797788288 & -4.56 & -8.16  & $\lambda$\,Ori  & [1]\\
  3340999023121894912 & -4.54 & -7.88  & $\lambda$\,Ori  & [1]\\
  5997006897751436544 & -4.70 & -9.32  & Lupus           & [2]\\
  5997082390415552768 & -4.73 & -8.73  & Lupus           & [2]\\
  6237190612936085120 & -4.74 & -7.80  & Upper Scorpius  & [3]\\
  6244083039015457664 & -4.58 & -9.05  & Upper Scorpius  & [3]\\
  6243393817024157184 & -4.64 & -10.51 & Upper Scorpius  & [3]\\
  6243130106031671168 & -4.75 & -10.97 & Upper Scorpius  & [3]\\
  6248768740952946816 & -4.61 & -9.58  & Upper Scorpius  & [3]\\
  6245777283349430912 & -4.53 & -8.76  & Upper Scorpius  & [3]\\
  5401795662560500352 & -4.37 & -8.59  & TW\,Hydra         & [4] \\
\hline\end{tabular}
\end{scriptsize}
\end{table}

Cross-match of  stars in the VHA-branch with the SIMBAD database shows that these objects are mostly classified as PMS stars, including YSO, T\,Tau, and Orion-type variables.
A few RS\,CVn's are also present.

In PMS stars, mass accretion can cause an excess flux in the \ion{Ca}{ii}\,IRT lines that dominates over the chromospheric flux (see Appendix\,\ref{sec:mass_accretion}).
Fig.\,\ref{fig:RpIRTvsMdot} shows the comparison of $R'_{\rm IRT}$ with mass accretion rate ($\log{\dot{M}}$) for 17 objects for which an estimate of the mass accretion rate is available.
The $\log{\dot{M}}$ estimates in Fig.\,\ref{fig:RpIRTvsMdot} comes from either the H$\alpha$ width at 10\% of the peak \citep{2004A&A...424..603N,2014A&A...561A...2A} measured by the Gaia-ESO survey \citep{2015A&A...576A..80L}\footnote{Data available at \url{https://www.eso.org/qi/catalogQuery/index/121}} or derived from the excess continuum emission  \cite[see, e.g.][]{2015A&A...579A..66M}.
The source identification, membership, and $\log{\dot{M}}$ references are reported in Table\,\ref{tab:accretion_ref}.
The $\log R'_{\rm IRT}$ vs $\log{\dot{M}}$ correlation is weak in all cases, i.e. considering the whole set, the measurements based on H$\alpha$ width at 10\% of the peak or on the excess continuum ($r \sim$ 0.3).
Nevertheless, all these objects have $R'_{\rm IRT} > -4.8$.

Previous knowledge \citep[e.g.][and references therein]{2017ARA&A..55..159L}, indicates that VHA stars on or close to the binary-MS can be identified as either chromospherically active dwarf stars in binary systems or PMS stars.
Stars located above the binary-MS can also be identified as PMS/T\,Tauri stars at an earlier phase of contraction towards the ZAMS, or, when they are located in the sub-giant region, as RS\,CVn systems in which at least one component is a sub-giant. 
The distribution of VHA stars in the CMD (Fig.\,\ref{fig:ai_CMD_gspphot}, upper panel), is consistent with this scenario, confirming the validity of the \gdr{3} activity index.
This situation suggests also that amongst stars in the very active branch ($R'_{\rm IRT} \gtrsim -4.8$) there is a superposition of mass-accreting stars and stars with enhanced activity, as that induced by tidal interaction in close binary system like RS\,CVn.
As a consequence, the separation  between the very high activity branch and the high activity branch seen in the upper left panel of Fig\,\ref{fig:RpIRTvsTeff} can  be due to the rapid decrease of mass accretion in PMS stars, but the very high activity branch is populated by both PMS accreting stars and other very active systems like RS\,CVn. 
The HA branch is therefore expected to be populated by chromospherically active stars with no mass accretion (see also Sect.\,\ref{sec:validation_rotmod}).
A dividing line between pure chromospheric activity and mass accretion dominated regimes can be placed at $\log R'_{\rm IRT}$ values between approximately -4.8 and -5.0, depending on \teff\ (see also Fig.\,\ref{fig:RpIRT_gspphot_histo}), but a value of $\log R'_{\rm IRT}$ above this limit may be due to enhanced activity in RS\,CVn systems as well.

\subsection{The High- and low-activity branches: the Vaughan-Preston gap}
\label{sec:VP}

\cite{1980PASP...92..385V} found a gap at some intermediate level of chromospheric activity in F and G stars observed in the Mount Wilson project.
This gap has been thoroughly discussed in the literature with different interpretations \citep[e.g. ][]{1981PASP...93..537D,1984ApJ...279..763N,1987A&A...177..131R,2009A&A...499L...9P}.
Recently, \cite{2018A&A...616A.108B} investigated this issue using their extensive compilation of $R'_{\rm HK}$, finding that for F, G, and early K stars this gap is indeed not so apparent.
They found, instead, that moving towards late-K and early-M dwarfs the concentration of stars on the active side of the gap increases and decreases on the low-activity side.

The plots in Fig.\,\ref{fig:RpIRTvsTeff} confirm the \cite{2018A&A...616A.108B} findings. 
Considering only stars with $\log g > 4.0$ and excluding the upper branch populated by pre-main sequence stars, we find no clear evidence of a gap at \teff\ corresponding to F, G, and early-K stars.
The middle panels of Fig.\,\ref{fig:RpIRTvsTeff}, on the other hand, confirm the disappearance of low-activity stars at decreasing \teff.
Below $\approx 4500$\,K, only the high activity branch essentially remains, which bends toward lower activity below $\approx 4000$\,K. 
Then, below $\approx 3500$\,K the spread in $R'_{\rm IRT}$ increases dramatically and the branch is not distinguishable any longer. 

\begin{figure*}
\begin{center}
\includegraphics[width=0.45\textwidth]{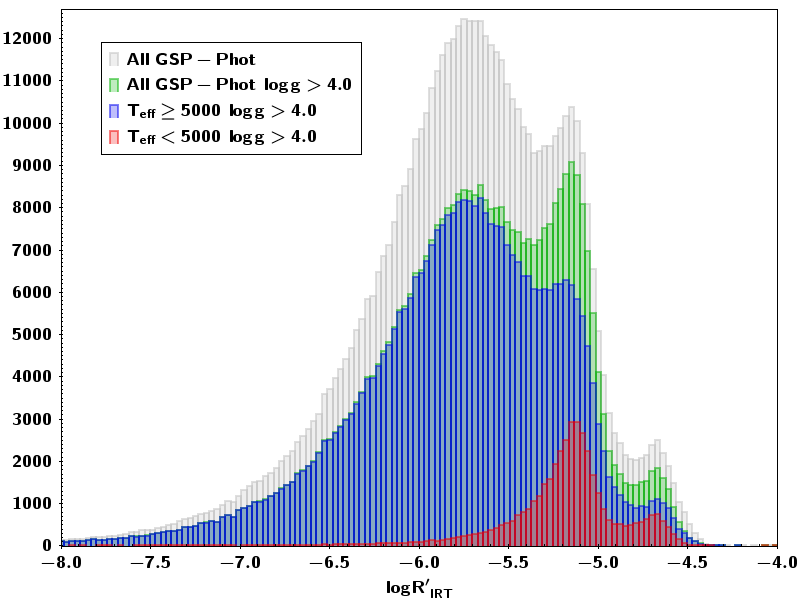}
\includegraphics[width=0.45\textwidth]{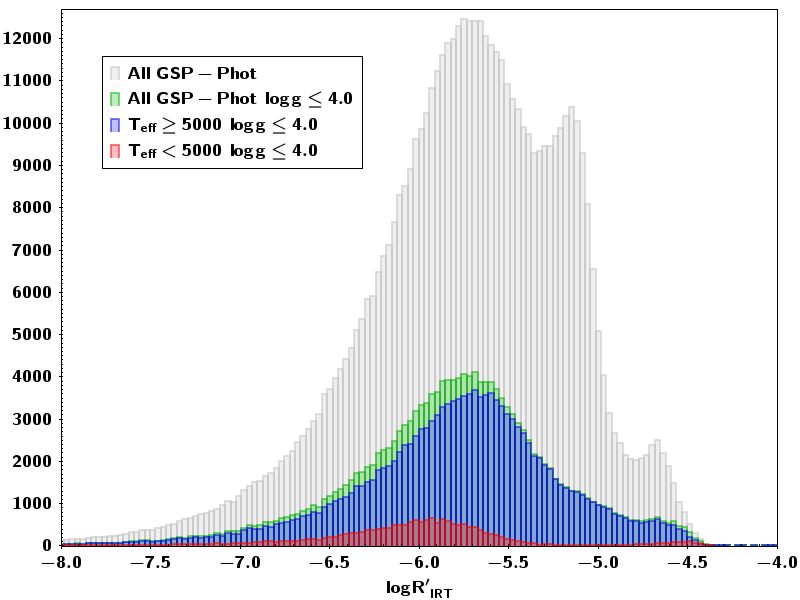}
\includegraphics[width=0.45\textwidth]{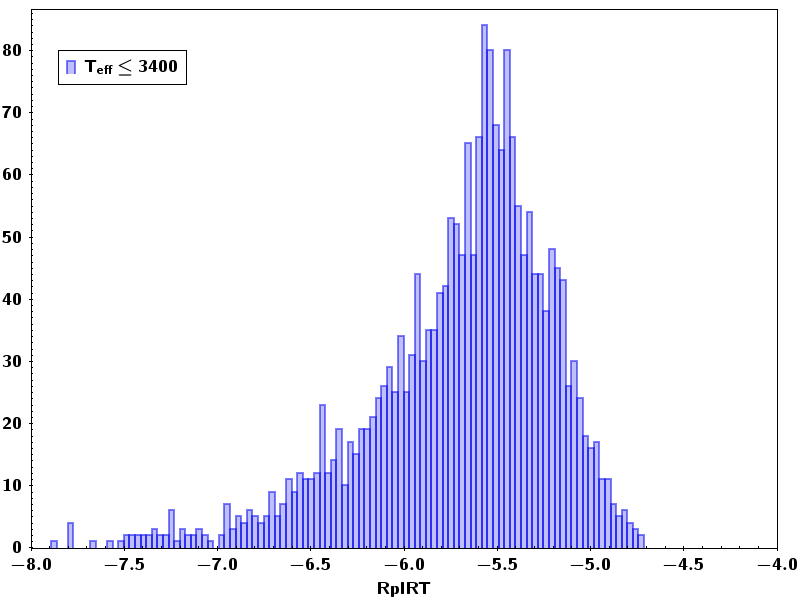}
\caption[]{$\log R'_{\rm IRT}$ histograms (bin size 0.03) for $\log g > 4.0$ (upper left panel) and $\log g \le 4.0$ (upper right panel). In the upper panels, the distribution for \teff $\ge 5000$\,K and \teff$< 5000$\,K are compared with the global distributions. 
The distribution for \teff$\apprle$\,3400\,K (fully convective stars) is shown in the bottom panel.
In order to extend the comparison to the very active branch on an homogeneous set, only results with \gspphot\ input are visualised.}
\label{fig:VP_gap}
\end{center}
\end{figure*}

Indeed, the histogram in Fig.\,\ref{fig:RpIRT_gspphot_histo} does show a decrease in the number of stars at $\log R'_{\rm IRT} \approx -5.4$, which would suggest a bimodal distribution in chromospheric activity with a clear minimum in between.
The histograms shown in Fig.\,\ref{fig:VP_gap}, however, shows that the two peaks on both side of this minimum are mostly due to the combination of the different $R'_{\rm IRT}$ distributions for stars cooler and hotter than  \teff $\approx$ 5000\,K.
Restricting the sample to stars with $\log g > 4.0$ and not considering stars in the highest activity branch populated by PMS stars, the distribution of stars with \teff $\gtrsim$ 5000\,K peaks at $\log R'_{\rm IRT} \approx -5.70$ and has only a bump at $R'_{\rm IRT} \approx -5.20$, showing only a shallow minimum in between, which remains so even restricting further the bin size.
On the other hand, the distribution of dwarfs with \teff $\lesssim$ 5000\,K peaks at $\log R'_{\rm IRT} \approx -5.13$ and shows no bump on neither sides of the maximum.
The cumulative distribution, which gathers together the different behaviour on either sides of \teff $\approx$ 5000\,K boundary, shows a double peak with a clear minimum at $\log R'_{\rm IRT} \approx -5.38$.

All stars with $\log g < 4.0$ (Fig.\,\ref{fig:VP_gap}, right panel) below $\log R'_{\rm IRT} \approx -4.8$ have a unimodal distribution, those with \teff$<$5000\,K peaking at $\log R'_{\rm IRT} \approx -5.9$ and those with \teff$<$5000\,K peaking at $\log R'_{\rm IRT} \approx -5.7$.
The distribution of stars with $\log g < 4.0$ has also a secondary peak above $\log R'_{\rm IRT} \approx -4.8$, which may be associated with sub-giants in close binaries with enhanced activity.

In summary, only dwarfs with \teff $\gtrsim$ 5000\,K show a hint of a bimodal distribution, with central values at $\log R'_{\rm IRT} \approx -5.70$ and -5.20, but without a clear gap in between.  
Notably, according to Eqs.\,\eqref{eq:RpIRTvsRpHKPhot} and \eqref{eq:RpIRTvsRpHKBS}, the Vaughan-Preston gap at $\log R'_{\rm HK} \approx -4.75$ corresponds to $\log R'_{\rm IRT} \approx -5.60$, which places it in between the estimated central values of the high (HA) and low-activity (LA) branches.

\subsection{Comparison with rotational modulation}
\label{sec:validation_rotmod}

\begin{figure*}
\begin{center}
\includegraphics[width=0.45\textwidth]{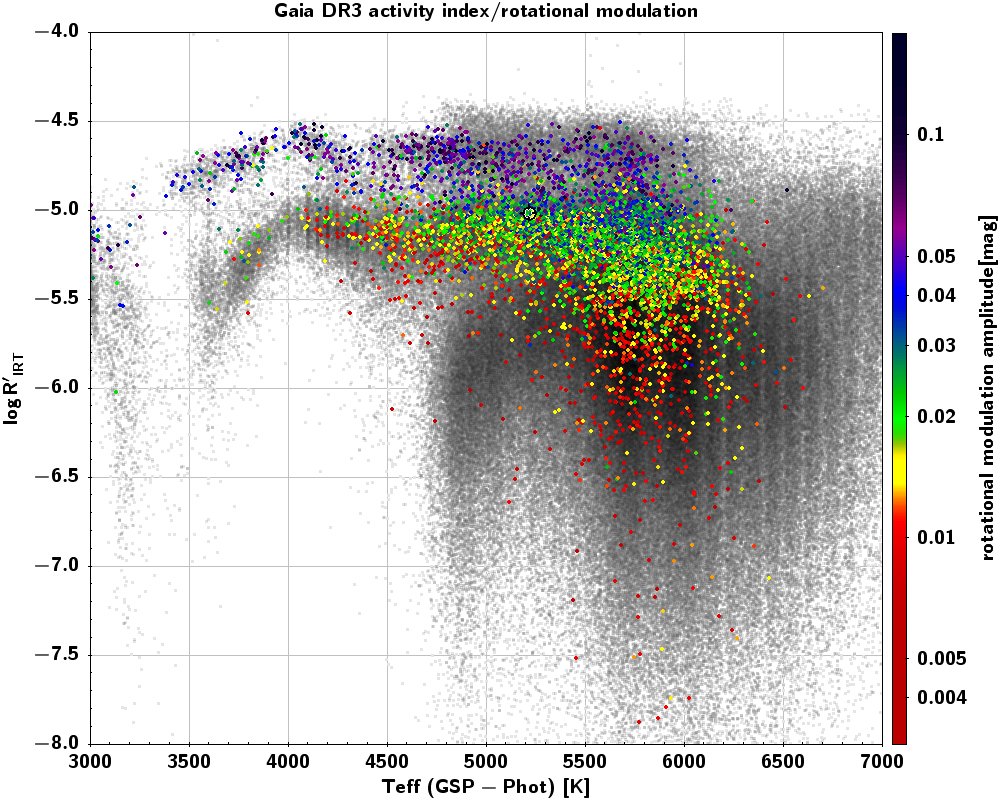}
\vspace{0.5cm}
\includegraphics[width=0.45\textwidth]{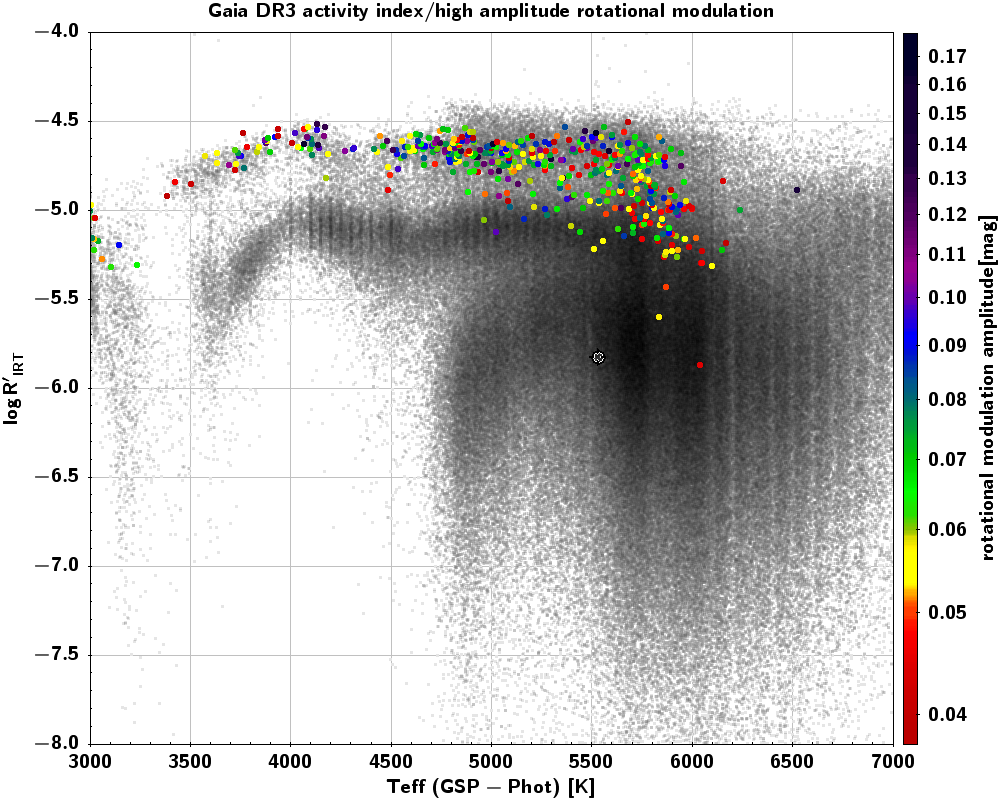}
\includegraphics[width=0.45\textwidth]{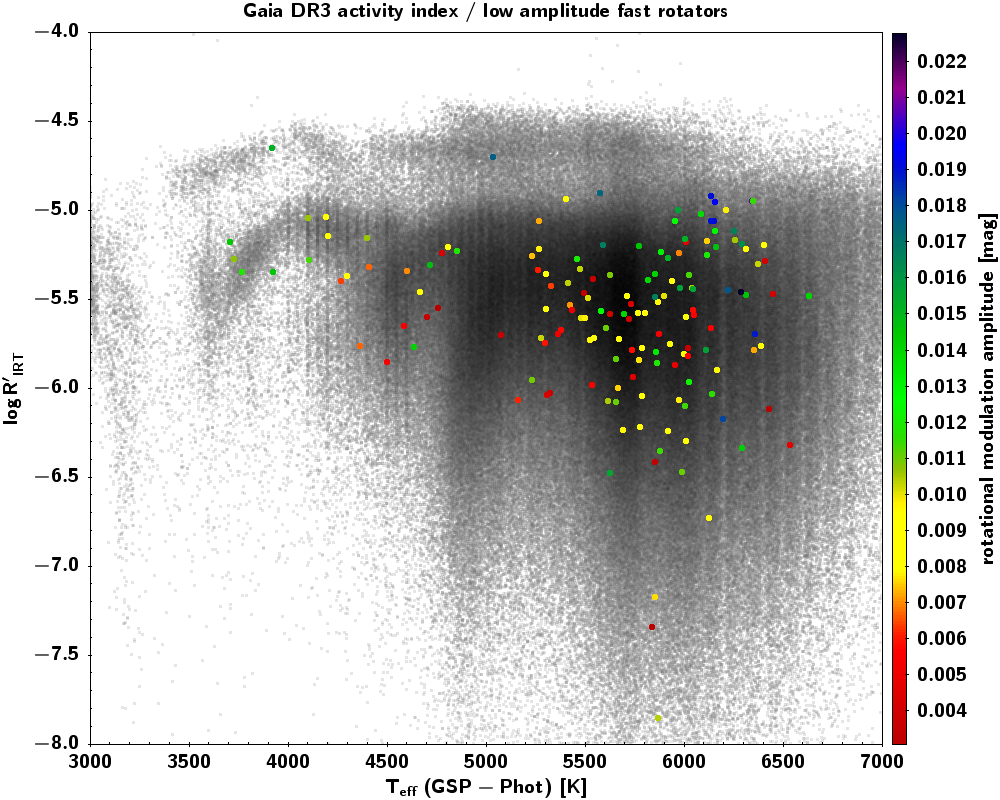}
\includegraphics[width=0.45\textwidth]{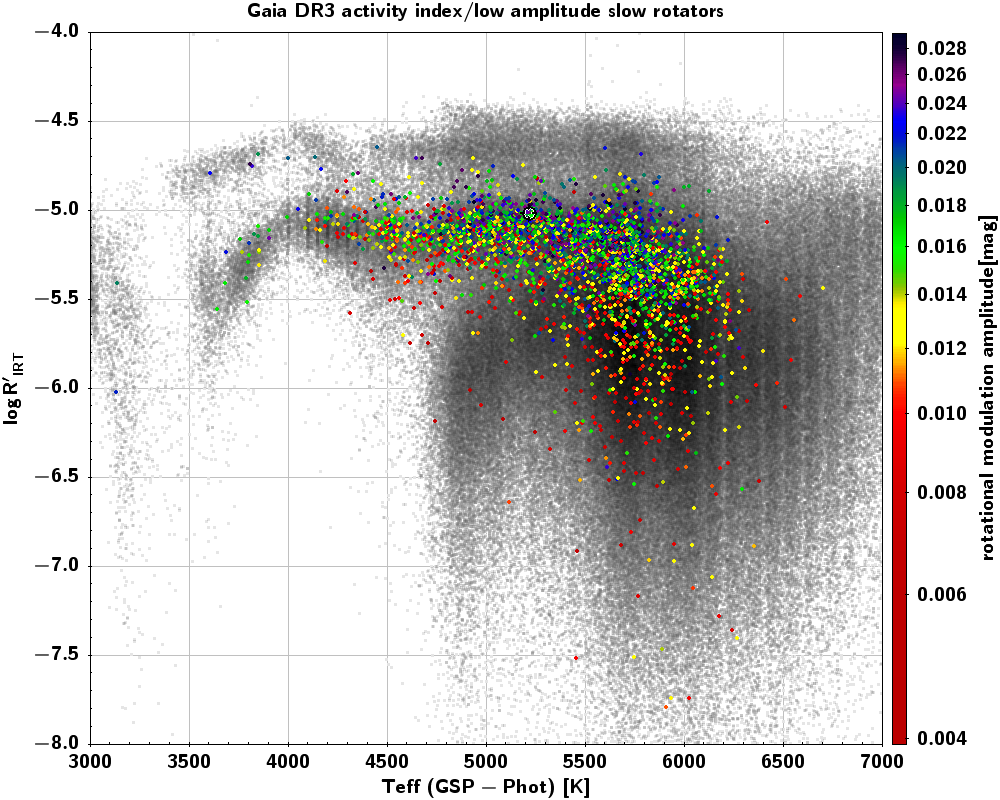}
\caption[]{$\log R'_{\rm IRT}$ vs \teff\ density diagram from (\teff from \gspphot) with \gdr{3}\ rotational modulation data overlaid (8171 sources in common). 
Top left panel: comparison with the whole rotational modulation dataset colour-coded according to the rotational modulation amplitude $A$. 
Top right panel: same as top left panel for just the high-amplitude rotator (HAR) branch.
Bottom left panel: same as top left panel for just the low-amplitude fast-rotator (LAFR) branch.
Bottom right panel: same as top left panel for just the low-amplitude slow-rotator (LASR) branch.
}
\label{fig:rotmod_vs_ai_gspphot}
\end{center}
\end{figure*}

The DPAC analysis includes the detection and characterisation of  rotational modulation due to photospheric active regions \citep[variability coordination unit, CU7,][]{2018A&A...616A..16L}.
\gdr{2} included rotational modulation data for some 150\,000 stars \citep{2018A&A...616A..16L} and \gdr{3} includes data for some 500\,000 stars \citep{DR3-DPACP-173}.

\cite{2019ApJ...877..157L} found that the rotational modulation amplitude distribution in \gdr{2} data shows a clear bimodality, with an evident gap at an amplitude $A \simeq 0.04 - 0.05$\,mag.
The low amplitude branch, in turn, shows a period bimodality with a main clustering at periods $P \approx$ 5 - 10\,d and a secondary clustering of ultra-fast rotators at $P \apprle 0.5$\,d.
These three different branches in the period--amplitude diagram were named high amplitude rotators (HAR), low-amplitude fast-rotators (LAFR), and low-amplitude slow-rotators (LASR).
\cite{2019ApJ...877..157L} interpreted these branches as signatures of different surface inhomogeneity regimes and suggested possible scenarios for their evolution.
The amplitude-period multimodality was found to be correlated with the position in the period-absolute magnitude (or period-color) diagram, with the low- and high-amplitude stars occupying different preferential locations.

Fig.\,\ref{fig:rotmod_vs_ai_gspphot} shows the positions of stars with different rotation modulation amplitude $A$ in the $\log R'_{\rm IRT}$ vs \teff\ density diagram.
A clear correlation between $R'_{\rm IRT}$ and $A$ is found ($r \approx 0.6$), as expected \citep[see e.g.][and references therein]{2017ARA&A..55..159L}, which increases when considered on a \teff\ by \teff\ basis ($r \approx 0.7$).
A confirmation on the consistency between rotational modulation and the activity index data in \gdr{3} comes also from the comparison with the branches identified in \cite{2019ApJ...877..157L}.
Stars in HAR-branch of the period-amplitude diagram are found to be located in the VHA-branch of the $\log R'_{\rm IRT}$ vs \teff\ density diagram.
Note that the HAR-branch is populated mainly by PMS stars \citep{2019ApJ...877..157L} as it is the VHA-branch (see Sect.\,\ref{sec:validation_accretion}).
Stars in the LASR-branch of the period--amplitude diagram are found in the HA and LA branches of the $\log R'_{\rm IRT}$ vs \teff\ density diagram, with the highest $A$ in this range mostly located in the HA branch.

The position in the $\log R'_{\rm IRT}$ vs \teff\ density diagram of the few LAFR stars in common is of particular interest.
\cite{2019ApJ...877..157L} suggested that the small modulation amplitude of stars in the LAFR branch may be due to a high degree of axisymmetry of the surface magnetic field. 
If this were the main difference compared to the HAR branch, one could naively expect the chromospheric activity in the LAFR and HAR branches to be quite similar.
The bottom left panel of Fig.\,\ref{fig:rotmod_vs_ai_gspphot} suggests, instead, a more complex scenario.
Most of the LAFR with \teff\,$\apprle$\,5000\,K are placed on the HA branch, the lowest available in this \teff\ range, with only one star on the VHA branch. 
LAFR stars with \teff\,$\apprge$\,5000\,K are mostly in the LA branch, with a few on the HA branch and only one in the VHA branch.
In general, there is no clear evidence of a correlation between chromospheric and photospheric activity in LAFR stars, but both are, unlike HAR stars, at low level.
This suggests that a high degree of axisymmetry of the surface magnetic field may not the only cause of the low rotational modulation amplitude in these stars, but rather that magnetic activity is inhibited in LAFR stars with respect to HAR stars with similar rotation period.
It may be argued that the LAFR sample may be contaminated by stars above the Kraft limit, i.e. stars with a radiative core and a convective envelope, because of an incorrect estimate of both insterstellar extinction and \teff.
The variability in this case could be due to stellar pulsation rather than rotational modulation.
Although we cannot exclude the presence of contaminants in the LAFR sample, there is evidence that at least a good fraction of them are real low-mass stars.
In fact, about 50 of them have APs in very good agreement between \gspphot\ and \gspspec, this latter being not affected by interstellar absorption.
Furthermore, from the stars' distance we estimate that for at least 50\% of them the interstellar extinction estimate by \gspphot\ should be biased by more than 3 mag/kpc to be high-mass stars that appear as low-mass stars, which is deemed quite unlikely.

\section{Conclusions}
\label{sec:conclusions}

\gdr{3} contains the largest chromospheric activity index catalogue obtained to date.
The activity index has been derived from the \ion{Ca}{ii}\,IRT observed by the \gaia\ Radial Velocity Spectrometer (RVS) for 2\,141\,643 stars in the Galaxy, with $G<13$, \teff$\in$(3000\,K, 7000\,K), \logg$\in$(3.0, 5.5), and \mh$\in$(-1.0, 1.0).
In this paper we described the method used and the results from the scientific validation.
We also provided a simple formula (Eq.\,\eqref{eq:conversion}) to convert the parameter published in the catalogue ({\fieldName{activityindex_espcs}}) to the $R'_{\rm{IRT}}$ index, analogous to the $R'_{\rm{HK}}$ index derived from the \ion{Ca}{ii} H \& K doublet, suitable for comparing the activity level over the whole range of parameters explored in \gdr{3}.

The scientific validation of the activity index obtained using the \gspphot\ parameters as input confirms the existence of three different regimes, which we called very high activity (VHA, $\log R'_{\rm{IRT}} \apprge -4.8$), high activity (HA, $-5.4 \apprle \log R'_{\rm{IRT}} \apprle -4.8$), and low activity (LA, $\log R'_{\rm{IRT}} \apprle -5.4$).
The activity index obtained using the \gspspec\ parameters as input do not show such clusterings because of an evident larger dispersion with respect to the results obtained with the \gspphot\ input and the \gspspec\ filtering applied to the most active stars like fast-rotator and PMS. 
The activity regimes found with the \gspphot\ input are consistent with those suggested by \cite{1996AJ....111..439H} and confirmed recently by \cite{2021A&A...646A..77G} from the analysis of $R'_{\rm{HK}}$.

The VHA regime is found to be populated by PMS stars and close binary system like RS\,CVn's. 
In the former, the excess flux with respect to the radiative equilibrium condition may be dominated by mass accretion.
In the latter, magnetic activity may be significantly enhanced by tidal interaction.
A comparison between $R'_{\rm IRT}$ and mass accretion rate $\dot{M}$ for a few PMS stars in the sample (Fig.\,\ref{fig:RpIRTvsMdot}) shows that, although the two parameters are not clearly correlated, a value $\log R'_{\rm IRT} \apprge -4.8$ may indicate a significant mass accretion flux contribution to the emission core of \ion{Ca}{ii}\,IRT.   
Indeed, the gap between the VHA branch and the rest of the distribution in the $\log R'_{\rm IRT}$ vs \teff\ diagram (Fig.\,\ref{fig:RpIRTvsTeff}) can be due to the rapid transition from the regime dominated by mass accretion to the regime dominated by magneto-acoustic heating of the chromosphere. 

The $R'_{\rm{IRT}}$ distribution (Fig.\,\ref{fig:VP_gap}) is found to depend on \teff\ and \logg.
Stars with \teff$<$\,5000\,K and \logg$>$\,4.0 are in the HA and VHA branches.
Excluding the VHA, the distribution of these stars in the HA branch is unimodal, peaking at $\log R'_{\rm{IRT}}\approx -5.13$.
The distribution of stars with \teff $\gtrsim$ 5000\,K and \logg$>$\,4.0, on the other hand, peaks at $\log R'_{\rm IRT} \approx -5.70$ and has a bump at $R'_{\rm IRT} \approx -5.20$, with a shallow minimum in between.
This minimum corresponds approximately to the location of the Vaughan-Preston gap at $\log R'_{\rm HK} \approx -4.75$ \citep{1980PASP...92..385V}, but this range is far from being void of stars, confirming \cite{2018A&A...616A.108B} finding.
Stars with \teff $\gtrsim$ 5000\,K and \logg$>$\,4.0 are therefore found in all three regimes, while stars with \teff $<$ 5000\,K and \logg$>$\,4.0 are found in the VHA and HA regimes only.

Stars with \logg$\in$(3.0,4.0) have unimodal distributions in the LA regime, peaking at $\log R'_{\rm IRT} \approx -5.9$ for \teff$<$5000\,K, and at $\log R'_{\rm IRT} \approx -5.7$ for \teff$\apprge$5000\,K.
There are also a few cases in the VHA regime, likely to be subgiants in close binaries with enhanced activity.

The activity distribution is found to change dramatically in the fully convective regime (Fig.\,\ref{fig:RpIRTvsTeff}). 
Stars close to the partially-convective / fully convective boundary have a large drop in chromospheric activity, then, at decreasing \teff\ (mass), activity increases again to reach approximately the same level of partially-convective stars in the HA branch.
A paucity of stars is found in the \teff -$\log R'_{\rm IRT}$ diagram at 3300\,K$\apprle$\teff$\apprle$\,3500\,K, likely connected to the gap in the mid-M dwarfs main sequence found by \cite{2018ApJ...861L..11J}, linked
to the onset of full convection in M dwarfs.

The $R'_{\rm IRT}$ index derived from the \gdr{3} activity index is well correlated with $R'_{\rm HK}$ from   \cite{2018A&A...616A.108B}, who gathered together a compilation of values from various surveys plus their own measurements on archive HARPS spectra.
We find also good correlation with a preliminary $R'_{\rm HK}$ estimate made on ESO-FEROS archive spectra, which will be subject of a future work.

The correlation of $R'_{\rm IRT}$ with the position in the \gaia\ CMD (Fig.\,\ref{fig:ai_CMD_gspphot}) is in line with the current knowledge of the activity evolution of low-mass stars \citep[see, e.g.][]{2017ARA&A..55..159L}.
Stars in the VHA branch, which according to a cross-match with the SIMBAD database includes PMS stars and RS\,CVn systems, are preferentially located above the single star MS, the higher $R'_{\rm IRT}$ the higher the tendency to be located at larger distance above the MS.
A group of very active early-K stars is found at $\sim$ 2--4 mag above the MS.
Some very active stars are also located close to the single- and binary-star MS.
Stars in the HA branch tend to be  on the single star MS, with some star on the binary-MS and a few above the MS.
The region of early-K stars at $\sim$ 2--4 mag above the MS is essentially void of HA stars.
Finally, stars in the LA branch are mostly concentrated in the F- G-type MS and in the sub-giant region, with the early-K region at $\sim$ 2--4 mag above the MS populated with some of the most inactive stars.
In this latter region we have therefore a superposition of VHA (close binaries or PMS stars) and LA stars (very inactive sub-giants).

The \teff -- $\log R'_{\rm IRT}$ diagram has a neat and unambiguous correspondence with the period--amplitude  diagram (PAD) of the rotational modulation due to photospheric spot and faculae \citep{2018A&A...616A..16L,2019ApJ...877..157L,DR3-DPACP-173}. 
Stars in the high amplitude rotator branch (HAR) of the PAD, as identified for the first time by \cite{2019ApJ...877..157L} in the \gdr{2} data, has a clear correspondence with the VHA branch in the \teff -- $\log R'_{\rm IRT}$ diagram.
Both are populated by young, PMS stars. 
The low-amplitude slow-rotator branch (LASR) of the PAD corresponds to the HA and LA branches of the \teff -- $\log R'_{\rm IRT}$ diagram.

The position of the low-amplitude fast-rotator (LAFR) stars in the \teff -- $\log R'_{\rm IRT}$ diagram is particularly interesting.
Despite the high rotation velocity, these stars have both low rotational modulation amplitude and most of them also low chromospheric activity index.
The former may be due either to a small photospheric active regions filling factor or to a high degree of axisymmetry of the surface magnetic field.
The latter can only be due to a global low level of magneto-acoustic heating of the chromosphere, with the axisymmetry of the surface magnetic field expected to  play essentially no role. 
The evidence points therefore to a global low level of photospheric and chromospheric activity rather than to a high degree of axisymmetry of the surface magnetic field in these fast rotating stars.
Moreover, \cite{DR3-DPACP-173}, who presented the rotational modulation data in \gdr{3}, find a low correlation between the \gmag\ variation due to rotational modulation and the \bmr\ colour variation in LAFR stars, which, combined with the low rotational modulation amplitude and the small chromospheric activity index, strengthen the evidence of the presence of smaller active regions in this branch compared with stars in the HAR branch with similar rotation period.
In general there is a good correlation between the chromospheric activity index and the amplitude of the rotational modulation due to spot and faculae, which is tighter when examined on limited colour or \teff\ bins.

Overall, the scientific validation of the \gdr{3} stellar activity index derived from the analysis of the \ion{Ca}{ii}\,IRT observed by RVS demonstrates a high level of consistency with previous results, most of them obtained from the analysis of the \ion{Ca}{ii} H \& K doublet, extending these results to a much larger sample.
The richness of the \gaia\ sample allows us to derive the distribution of the stellar activity over stellar fundamental parameters with an unprecedented level of details, outlining different regimes of chromospheric heating and identifying those where the emission due to mass accretion processes may dominate over a purely chromospheric emission.
The activity index provided in \gdr{3} represents therefore a gold mine for several investigations like: verifying/falsifying theoretical models of the stellar dynamo and of the emergence of stellar active regions; identifying stars with a low RV jitter in the search of Earth-like planets; studying the impact of magnetic activity in the derivation of accurate elemental abundance in stellar atmospheres; studying the evolution of the magnetic activity for different stellar mass and its impact on the host exoplanets; studying the magnetic activity in close binary systems.


\appendix

\section{Chromospheric activity indices}\label{sec:activity_indices}

A chromospheric activity index gives a relative measure of the non-radiative energy dissipated in the stellar chromosphere.
For thorough reviews of our current understanding of stellar chromospheres we refer to \cite{2017ARA&A..55..159L} and \cite{2019ARA&A..57..189C}.
Here, we summarise the basic concepts at the basis of the definition of a chromospheric activity index, its usage as chromospheric diagnostics, and compare the activity index defined in this work with other indices and definitions in the literature.

The best known chromospheric activity index is the \rprimeHK\ derived from the Ca\,II H and K doublet at $\lambda = 393.4$\,nm and 396.9\,nm \citep[see, e.g.][ and references therein]{1979ApJS...41...47L,1984ApJ...279..763N}.
This index provides a measure of the chromospheric radiative flux in the H and K lines, in units of the bolometric flux:
\begin{equation}
\label{eq:rp_definition}
R'_{\rm{HK}} \equiv \frac{\mathcal{F'}(\rm{H}_1)+\mathcal{F'}(\rm{K}_1)}{\sigma T_{\rm eff}^4}
\end{equation}
where $\mathcal{F'}(\rm{H}_1)$ and $\mathcal{F'}(\rm{K}_1)$ are the chromospheric contribution to the flux at stellar surface in the H and K lines and $\sigma$ the Boltzmann constant.
The chromospheric flux in the Ca\,II H and K doublet is identified as the flux in the emission core of the lines, integrated over the wavelength ranges $\rm{H}_1$ and $\rm{K}_1$ between the flux minima, corrected by the photospheric contribution: 
\begin{eqnarray}
\mathcal{F'}(\rm{H}_1) =  \mathcal{F}(\rm{H}_1) - \mathcal{F}_{\rm phot}(\rm{H}_1) \\
\mathcal{F'}(\rm{K}_1) =  \mathcal{F}(\rm{K}_1) - \mathcal{F}_{\rm phot}(\rm{K}_1)
\end{eqnarray}
  
A purely empirical derivation of \rprimeHK\ requires narrow-band absolute photometry \citep{1978PASP...90..267V} or a combination of spectroscopy and narrow-band absolute photometry \citep{1979ApJS...41...47L}.
In these cases,  the calibration of $\mathcal{F}(\rm{H}_1)$ and $\mathcal{F}(\rm{K}_1)$ is performed by taking the ratio of the signal in the core of the lines and in nearby wavelength intervals \citep[e.g. the $S$ index, ][]{1978PASP...90..267V,1984ApJ...279..763N} and then applying an absolute flux calibration to the nearby intervals.
A purely empirical derivation of \rprimeHK\ also requires knowledge of the angular stellar diameter $\phi'$ since, ignoring interstellar absorption, the flux at stellar surface $\mathcal{F}(\Delta \lambda)$ is related to the observed flux $f(\Delta \lambda)$ by:
\begin{equation}
\mathcal{F}(\Delta \lambda) = f(\Delta \lambda) \left(\frac{d}{R}\right)^2 = f(\Delta \lambda) \left(\frac{4.125 \cdot 10^8}{\phi'}\right)^2
\end{equation}
with $\phi'$ in mas. 

The low NUV continuum flux in late-type stars makes diagnostics like Ca\,II H and K and Mg h and k (at $\lambda= 279.6$\,nm and 280.4\,nm) particularly sensitive to chromospheric activity, as the core of these lines may be driven into emission more easily than in spectral regions where the photospheric continuum flux is much higher.
On the other hand, a faint underlying continuum makes the flux calibration technique described above difficult, and therefore this approach may become unfeasible below some limiting \teff.

A similar approach can be applied to other chromospheric diagnostics having a much higher underlying continuum, like H$\alpha$ and \ion{Ca}{ii}\,IRT.
In such cases, because of the unfavourable line/continuum contrast effect, low or moderate chromospheric activity produce some extra absorption \citep{1979ApJ...234..579C} or just a fill-in in the line core; only relatively high levels of magnetic activity may drive the core of the lines into emission.

The chromospheric component of the line can be revealed by comparing the target spectrum with a purely photospheric spectrum with the same \teff, \logg, and \mh.
When a non-negligible chromospheric activity is present, the difference or ratio spectrum reveals some extra absorption or emission.
An estimate of the chromospheric flux can be conveniently done using the residual equivalent width
\begin{equation}
\label{eq:def_residual_ew}
\Delta W \equiv \bigintssss_{\Delta \lambda} \left[r(\lambda) - r_{\rm phot}(\lambda)\right]\, d\lambda\,, 
\end{equation}
where
\begin{equation}
r(\lambda) \equiv \frac{f(\lambda)}{f_c (\lambda)} 
\end{equation}
is the continuum normalised spectrum, so that
\begin{equation}
\label{eq:chromoflux}
\mathcal{F'}(\Delta \lambda) = \mathcal{F}_c \Delta W
\end{equation}
where $\mathcal{F}_c$ is the absolute flux in the underlying or nearby continuum.
With the help of Eq.\,\eqref{eq:chromoflux}, the definition of \rprimeHK can be generalised to any spectral line whose core is formed in the chromosphere:
\begin{equation}
\label{eq:rp_gen_definition}
R'_{\rm{\Delta \lambda}} \equiv \frac{\mathcal{F'}(\Delta \lambda)}{\sigma T_{\rm eff}^4}
\end{equation}
where $\Delta \lambda$ is the approximate wavelength range embracing the line core.
Because of the increasing line absorption for increasing activity levels in the low activity regime, the relationship between $\Delta W$ and the activity level is expected to be double-valued for given \teff, \logg, and \mh, and therefore only activity levels above a certain threshold can be put on a relative scale using lines with a strong underlying continuum like H$\alpha$ and \ion{Ca}{ii}\,IRT.

The photospheric reference spectrum $r_{\rm phot}(\lambda)$ can be taken as the spectrum of an inactive star with \teff, \logg, and \mh\ as close as possible to the target star \citep{1985ApJ...289..269H}, a combination of inactive star spectra \citep[see, e.g.,][]{1994A&A...284..883F,2015A&A...575A...4F,2015A&A...576A..80L}, or synthetic spectra \citep[e.g.][and references therein]{2005A&A...430..669A,2007A&A...466.1089B}.

The choice of adopting observed spectra of inactive stars as  template $r_{\rm phot}$ in Eq.\,\eqref{eq:def_residual_ew} aims at obtaining results that are as much as possible model independent.
The main conceptual problem here is how to define an inactive star. From a purely observational point of view, the seemingly obvious choice is to select stars with the deepest absorption amongst those with similar \teff, \logg, and \mh\ \citep[e.g.][]{2015A&A...576A..80L,2015A&A...575A...4F,2017ApJ...835...61Z}.
As noted above, this introduces ambiguities, since in the presence of a strong underlying continuum the deepest absorption does not correspond to the lowest chromospheric activity level. 
Another important caveat is that setting up a grid of reference inactive stellar spectra that cover the relevant parameter space with sufficient density and accuracy can be costly and prone to subjective choices.

It should also be noted that photospheric synthetic spectra are built under the assumption of radiative equilibrium (RE), but chromospheric activity is superimposed on a basal emission originating from the dissipation of acoustic energy \citep{1989ApJ...337..964S}, and therefore it is expected that RE breaks down in the upper photosphere of all late-type stars, even for the lowest chromospheric activity level.
When synthetic spectra are adopted as templates, $r_{\rm phot}$ in Eq.\,\eqref{eq:def_residual_ew} is replaced with $r_{\rm RE}$, the continuum normalised spectrum under the RE assumption.
Since this is deduced from simple {\it ab initio} assumption, $r_{\rm RE}$ is a robust reference, but it is not expected to be realised in the upper photosphere of any real star.
Non-local thermodynamic equilibrium (NLTE) effects in the line core should also be taken into account. 
However, \cite{2005A&A...430..669A} found that NLTE effect on the \ion{Ca}{ii}\,IRT lines equivalent widths are negligible (10\%) except for very low-metallicity models (\mh $\simeq$ -2.0, 30\%), increasing with decreasing \teff\ and decreasing \logg.

In summary, using templates built by interpolating over a grid of LTE synthetic spectra is a practical and robust choice by which stars can be put on a relative chromospheric activity scale provided:
\begin{enumerate}[label=\alph*)]
\item $\Delta W$ is positive and above a certain threshold set by uncertainties in the model and observations;
\item A scale of $\Delta W$ is built for similar values of \teff, \logg, and \mh, or, alternatively, a relative flux scale is built using Eq.\,\eqref{eq:rp_gen_definition}
\end{enumerate}


\section{\ion{Ca}{ii}\,IRT as mass accretion indicator}\label{sec:mass_accretion}

In the pre-main sequence phase, after the main phase of mass accretion, young stars, both classical T\,Tauri stars (CTTSs) and the higher mass Herbig AeBe stars, accrete mass from their protostellar disk. 
A generally accepted model for this evolutionary stage \citep[magnetospheric infall and accretion shock, ][]{2016ARA&A..54..135H,1998ApJ...509..802C,2001ApJ...550..944M} involves downflows from the protostellar disk along magnetic field lines which produce a shock near the stellar photosphere.
X-ray and UV photons generated in the shock front both heat the photosphere below, producing hot spots, and are reprocessed in the accreting gas to produce far-UV (FUV) and near-UV (NUV) continua and emission lines.
This model explains the observed hot optical/UV excess continuum emission, which manifest itself in the Balmer jump emission and the filling in, commonly referred to as ``veiling'', of photospheric absorption lines; its strength is correlated with the mass infall rate \citep[e.g.][and references therein]{1998AJ....116..455M}.
Key features of emission lines in CTTSs include redshifted absorption components, blue-ward asymmetries in the profile, as well as the presence of broad and narrow components.
The narrow component observed in the \ion{Ca}{ii}\,IRT and \ion{He}{i}\,(587.6\,nm) was found to be correlated with veiling and NIR excess by \cite{1996ApJS..103..211B}.
\cite{1998AJ....116..455M}, on the other hand, suggested that the broad emission component of the H$\alpha$, the higher Paschen series, \ion{Ca}{ii}\,IRT, \ion{He}{i}\,(587.6\,nm), \ion{O}{i}\,(777.3,844.6\,nm), and \ion{Na}{}\,D are formed in the turbulent magnetosphere of the star.
Many weak line T\,Tauri stars (WTTS) have the same age of CTTSs but do not show excess continuum emission nor strong emission lines.
This support the scenario in which once accretion is completed the CTTSs become WTTSs; in CTTSs the continuum excess emission and the strong emission lines are dominated by the accretion processes, while only magnetic activity characterise the WTTSs spectra. 
The magnetic activity and accretion effects are superimposed, and while high mass accretion rates produce continuum and lines emission that dominate completely the stellar chromospheric emission, for the mostly slowly accreting stars the stellar chromosphere may hide the accretion emission \citep{2011ApJ...743..105I}.

High level of the \ion{Ca}{ii}\,IRT activity index are therefore undoubtedly associated with mass accretion in CTTSs.
However, the transition from a chromospheric- to accretion-dominated regimes is not sharp and the two regimes are expected to overlap somewhat. 
In the context of the RAVE survey, \cite{2017ApJ...835...61Z} found a bimodal distribution of the excess emission in the \ion{Ca}{ii}\,IRT with a primary peak associated with ``inactive'' stars and a secondary peak at higher excess emission associated with young stars.
A much well defined boundary between chromospheric activity and accretion dominated regimes has been found by \cite{2015A&A...576A..80L,2015A&A...575A...4F,2003AJ....126.2997B} analysing the H$\alpha$ emission flux as a function of \teff.


\section*{Acknowledgements}
This work presents results from the European Space Agency (ESA) space mission \gaia. \gaia\ data are processed by the \gaia\ Data Processing and Analysis Consortium (DPAC). Funding for the DPAC is provided by national institutions, in particular the institutions participating in the \gaia\ MultiLateral Agreement (MLA). The \gaia\ mission website is \url{https://www.cosmos.esa.int/gaia}. The \gaia\ archive website is \url{https://archives.esac.esa.int/gaia}.

The \gaia\ mission and data processing have financially been supported by, in alphabetical order by country:
\begin{itemize}
\item the Algerian Centre de Recherche en Astronomie, Astrophysique et G\'{e}ophysique of Bouzareah Observatory;
\item the Austrian Fonds zur F\"{o}rderung der wissenschaftlichen Forschung (FWF) Hertha Firnberg Programme through grants T359, P20046, and P23737;
\item the BELgian federal Science Policy Office (BELSPO) through various PROgramme de D\'{e}veloppement d'Exp\'{e}riences scientifiques (PRODEX) grants and the Polish Academy of Sciences - Fonds Wetenschappelijk Onderzoek through grant VS.091.16N, and the Fonds de la Recherche Scientifique (FNRS), and the Research Council of Katholieke Universiteit (KU) Leuven through grant C16/18/005 (Pushing AsteRoseismology to the next level with TESS, GaiA, and the Sloan DIgital Sky SurvEy -- PARADISE);  
\item the Brazil-France exchange programmes Funda\c{c}\~{a}o de Amparo \`{a} Pesquisa do Estado de S\~{a}o Paulo (FAPESP) and Coordena\c{c}\~{a}o de Aperfeicoamento de Pessoal de N\'{\i}vel Superior (CAPES) - Comit\'{e} Fran\c{c}ais d'Evaluation de la Coop\'{e}ration Universitaire et Scientifique avec le Br\'{e}sil (COFECUB);
\item the Chilean Agencia Nacional de Investigaci\'{o}n y Desarrollo (ANID) through Fondo Nacional de Desarrollo Cient\'{\i}fico y Tecnol\'{o}gico (FONDECYT) Regular Project 1210992 (L.~Chemin);
\item the National Natural Science Foundation of China (NSFC) through grants 11573054, 11703065, and 12173069, the China Scholarship Council through grant 201806040200, and the Natural Science Foundation of Shanghai through grant 21ZR1474100;  
\item the Tenure Track Pilot Programme of the Croatian Science Foundation and the \'{E}cole Polytechnique F\'{e}d\'{e}rale de Lausanne and the project TTP-2018-07-1171 `Mining the Variable Sky', with the funds of the Croatian-Swiss Research Programme;
\item the Czech-Republic Ministry of Education, Youth, and Sports through grant LG 15010 and INTER-EXCELLENCE grant LTAUSA18093, and the Czech Space Office through ESA PECS contract 98058;
\item the Danish Ministry of Science;
\item the Estonian Ministry of Education and Research through grant IUT40-1;
\item the European Commission?s Sixth Framework Programme through the European Leadership in Space Astrometry (\href{https://www.cosmos.esa.int/web/gaia/elsa-rtn-programme}{ELSA}) Marie Curie Research Training Network (MRTN-CT-2006-033481), through Marie Curie project PIOF-GA-2009-255267 (Space AsteroSeismology \& RR Lyrae stars, SAS-RRL), and through a Marie Curie Transfer-of-Knowledge (ToK) fellowship (MTKD-CT-2004-014188); the European Commission's Seventh Framework Programme through grant FP7-606740 (FP7-SPACE-2013-1) for the \gaia\ European Network for Improved data User Services (\href{https://gaia.ub.edu/twiki/do/view/GENIUS/}{GENIUS}) and through grant 264895 for the \gaia\ Research for European Astronomy Training (\href{https://www.cosmos.esa.int/web/gaia/great-programme}{GREAT-ITN}) network;
\item the European Cooperation in Science and Technology (COST) through COST Action CA18104 `Revealing the Milky Way with \gaia\ (MW-Gaia)';
\item the European Research Council (ERC) through grants 320360, 647208, and 834148 and through the European Union's Horizon 2020 research and innovation and excellent science programmes through Marie Sk{\l}odowska-Curie grant 745617 (Our Galaxy at full HD -- Gal-HD) and 895174 (The build-up and fate of self-gravitating systems in the Universe) as well as grants 687378 (Small Bodies: Near and Far), 682115 (Using the Magellanic Clouds to Understand the Interaction of Galaxies), 695099 (A sub-percent distance scale from binaries and Cepheids -- CepBin), 716155 (Structured ACCREtion Disks -- SACCRED), 951549 (Sub-percent calibration of the extragalactic distance scale in the era of big surveys -- UniverScale), and 101004214 (Innovative Scientific Data Exploration and Exploitation Applications for Space Sciences -- EXPLORE);
\item the European Science Foundation (ESF), in the framework of the \gaia\ Research for European Astronomy Training Research Network Programme (\href{https://www.cosmos.esa.int/web/gaia/great-programme}{GREAT-ESF});
\item the European Space Agency (ESA) in the framework of the \gaia\ project, through the Plan for European Cooperating States (PECS) programme through contracts C98090 and 4000106398/12/NL/KML for Hungary, through contract 4000115263/15/NL/IB for Germany, and through PROgramme de D\'{e}veloppement d'Exp\'{e}riences scientifiques (PRODEX) grant 4000127986 for Slovenia;  
\item the Academy of Finland through grants 299543, 307157, 325805, 328654, 336546, and 345115 and the Magnus Ehrnrooth Foundation;
\item the French Centre National d'\'{E}tudes Spatiales (CNES), the Agence Nationale de la Recherche (ANR) through grant ANR-10-IDEX-0001-02 for the `Investissements d'avenir' programme, through grant ANR-15-CE31-0007 for project `Modelling the Milky Way in the \gaia era' (MOD4Gaia), through grant ANR-14-CE33-0014-01 for project `The Milky Way disc formation in the \gaia era' (ARCHEOGAL), through grant ANR-15-CE31-0012-01 for project `Unlocking the potential of Cepheids as primary distance calibrators' (UnlockCepheids), through grant ANR-19-CE31-0017 for project `Secular evolution of galxies' (SEGAL), and through grant ANR-18-CE31-0006 for project `Galactic Dark Matter' (GaDaMa), the Centre National de la Recherche Scientifique (CNRS) and its SNO \gaia of the Institut des Sciences de l'Univers (INSU), its Programmes Nationaux: Cosmologie et Galaxies (PNCG), Gravitation R\'{e}f\'{e}rences Astronomie M\'{e}trologie (PNGRAM), Plan\'{e}tologie (PNP), Physique et Chimie du Milieu Interstellaire (PCMI), and Physique Stellaire (PNPS), the `Action F\'{e}d\'{e}ratrice \gaia' of the Observatoire de Paris, the R\'{e}gion de Franche-Comt\'{e}, the Institut National Polytechnique (INP) and the Institut National de Physique nucl\'{e}aire et de Physique des Particules (IN2P3) co-funded by CNES;
\item the German Aerospace Agency (Deutsches Zentrum f\"{u}r Luft- und Raumfahrt e.V., DLR) through grants 50QG0501, 50QG0601, 50QG0602, 50QG0701, 50QG0901, 50QG1001, 50QG1101, 50\-QG1401, 50QG1402, 50QG1403, 50QG1404, 50QG1904, 50QG2101, 50QG2102, and 50QG2202, and the Centre for Information Services and High Performance Computing (ZIH) at the Technische Universit\"{a}t Dresden for generous allocations of computer time;
\item the Hungarian Academy of Sciences through the Lend\"{u}let Programme grants LP2014-17 and LP2018-7 and the Hungarian National Research, Development, and Innovation Office (NKFIH) through grant KKP-137523 (`SeismoLab');
\item the Science Foundation Ireland (SFI) through a Royal Society - SFI University Research Fellowship (M.~Fraser);
\item the Israel Ministry of Science and Technology through grant 3-18143 and the Tel Aviv University Center for Artificial Intelligence and Data Science (TAD) through a grant;
\item the Agenzia Spaziale Italiana (ASI) through contracts I/037/08/0, I/058/10/0, 2014-025-R.0, 2014-025-R.1.2015, and 2018-24-HH.0 to the Italian Istituto Nazionale di Astrofisica (INAF), contract 2014-049-R.0/1/2 to INAF for the Space Science Data Centre (SSDC, formerly known as the ASI Science Data Center, ASDC), contracts I/008/10/0, 2013/030/I.0, 2013-030-I.0.1-2015, and 2016-17-I.0 to the Aerospace Logistics Technology Engineering Company (ALTEC S.p.A.), INAF, and the Italian Ministry of Education, University, and Research (Ministero dell'Istruzione, dell'Universit\`{a} e della Ricerca) through the Premiale project `MIning The Cosmos Big Data and Innovative Italian Technology for Frontier Astrophysics and Cosmology' (MITiC);
\item the Netherlands Organisation for Scientific Research (NWO) through grant NWO-M-614.061.414, through a VICI grant (A.~Helmi), and through a Spinoza prize (A.~Helmi), and the Netherlands Research School for Astronomy (NOVA);
\item the Polish National Science Centre through HARMONIA grant 2018/30/M/ST9/00311 and DAINA grant 2017/27/L/ST9/03221 and the Ministry of Science and Higher Education (MNiSW) through grant DIR/WK/2018/12;
\item the Portuguese Funda\c{c}\~{a}o para a Ci\^{e}ncia e a Tecnologia (FCT) through national funds, grants SFRH/\-BD/128840/2017 and PTDC/FIS-AST/30389/2017, and work contract DL 57/2016/CP1364/CT0006, the Fundo Europeu de Desenvolvimento Regional (FEDER) through grant POCI-01-0145-FEDER-030389 and its Programa Operacional Competitividade e Internacionaliza\c{c}\~{a}o (COMPETE2020) through grants UIDB/04434/2020 and UIDP/04434/2020, and the Strategic Programme UIDB/\-00099/2020 for the Centro de Astrof\'{\i}sica e Gravita\c{c}\~{a}o (CENTRA);  
\item the Slovenian Research Agency through grant P1-0188;
\item the Spanish Ministry of Economy (MINECO/FEDER, UE), the Spanish Ministry of Science and Innovation (MICIN), the Spanish Ministry of Education, Culture, and Sports, and the Spanish Government through grants BES-2016-078499, BES-2017-083126, BES-C-2017-0085, ESP2016-80079-C2-1-R, ESP2016-80079-C2-2-R, FPU16/03827, PDC2021-121059-C22, RTI2018-095076-B-C22, and TIN2015-65316-P (`Computaci\'{o}n de Altas Prestaciones VII'), the Juan de la Cierva Incorporaci\'{o}n Programme (FJCI-2015-2671 and IJC2019-04862-I for F.~Anders), the Severo Ochoa Centre of Excellence Programme (SEV2015-0493), and MICIN/AEI/10.13039/501100011033 (and the European Union through European Regional Development Fund `A way of making Europe') through grant RTI2018-095076-B-C21, the Institute of Cosmos Sciences University of Barcelona (ICCUB, Unidad de Excelencia `Mar\'{\i}a de Maeztu') through grant CEX2019-000918-M, the University of Barcelona's official doctoral programme for the development of an R+D+i project through an Ajuts de Personal Investigador en Formaci\'{o} (APIF) grant, the Spanish Virtual Observatory through project AyA2017-84089, the Galician Regional Government, Xunta de Galicia, through grants ED431B-2021/36, ED481A-2019/155, and ED481A-2021/296, the Centro de Investigaci\'{o}n en Tecnolog\'{\i}as de la Informaci\'{o}n y las Comunicaciones (CITIC), funded by the Xunta de Galicia and the European Union (European Regional Development Fund -- Galicia 2014-2020 Programme), through grant ED431G-2019/01, the Red Espa\~{n}ola de Supercomputaci\'{o}n (RES) computer resources at MareNostrum, the Barcelona Supercomputing Centre - Centro Nacional de Supercomputaci\'{o}n (BSC-CNS) through activities AECT-2017-2-0002, AECT-2017-3-0006, AECT-2018-1-0017, AECT-2018-2-0013, AECT-2018-3-0011, AECT-2019-1-0010, AECT-2019-2-0014, AECT-2019-3-0003, AECT-2020-1-0004, and DATA-2020-1-0010, the Departament d'Innovaci\'{o}, Universitats i Empresa de la Generalitat de Catalunya through grant 2014-SGR-1051 for project `Models de Programaci\'{o} i Entorns d'Execuci\'{o} Parallels' (MPEXPAR), and Ramon y Cajal Fellowship RYC2018-025968-I funded by MICIN/AEI/10.13039/501100011033 and the European Science Foundation (`Investing in your future');
\item the Swedish National Space Agency (SNSA/Rymdstyrelsen);
\item the Swiss State Secretariat for Education, Research, and Innovation through the Swiss Activit\'{e}s Nationales Compl\'{e}mentaires and the Swiss National Science Foundation through an Eccellenza Professorial Fellowship (award PCEFP2\_194638 for R.~Anderson);
\item the United Kingdom Particle Physics and Astronomy Research Council (PPARC), the United Kingdom Science and Technology Facilities Council (STFC), and the United Kingdom Space Agency (UKSA) through the following grants to the University of Bristol, the University of Cambridge, the University of Edinburgh, the University of Leicester, the Mullard Space Sciences Laboratory of University College London, and the United Kingdom Rutherford Appleton Laboratory (RAL): PP/D006511/1, PP/D006546/1, PP/D006570/1, ST/I000852/1, ST/J005045/1, ST/K00056X/1, ST/\-K000209/1, ST/K000756/1, ST/L006561/1, ST/N000595/1, ST/N000641/1, ST/N000978/1, ST/\-N001117/1, ST/S000089/1, ST/S000976/1, ST/S000984/1, ST/S001123/1, ST/S001948/1, ST/\-S001980/1, ST/S002103/1, ST/V000969/1, ST/W002469/1, ST/W002493/1, ST/W002671/1, ST/W002809/1, and EP/V520342/1.
\end{itemize}

We acknowledge the use of TOPCAT \citep[\href{http://www.starlink.ac.uk/topcat/}{http://www.starlink.ac.uk/topcat/},][]{2005ASPC..347...29T}.

\bibliographystyle{aa} 
\bibliography{Apsis_ESP_CS}

\end{document}